\documentclass[journal,twocolumn]{IEEEtran}
\IEEEoverridecommandlockouts
\usepackage{color}
\usepackage{leftidx}
\usepackage{cite}
\usepackage{multirow}
\usepackage{subfigure}
\usepackage{graphicx,amssymb,amstext,amsmath}
\usepackage{cleveref}
\usepackage[ruled,vlined,lined,ruled,linesnumbered]{algorithm2e}
\usepackage{url}
\usepackage[font=small,skip=0pt]{caption}
\usepackage[belowskip=-15pt,aboveskip=10pt]{caption}

\allowdisplaybreaks
\begin{document}

\title{Lyapunov-guided Deep Reinforcement Learning for Stable Online Computation Offloading in Mobile-Edge Computing Networks}

\author{Suzhi~Bi, ~\IEEEmembership{Senior Member,~IEEE}, Liang~Huang, ~\IEEEmembership{Member,~IEEE}, Hui Wang, and Ying-Jun Angela Zhang, ~\IEEEmembership{Fellow,~IEEE}\\
\thanks{S.~Bi is with the College of Electronics and Information Engineering, Shenzhen University, Shenzhen, China 518060 (e-mail: bsz@szu.edu.cn). S.~Bi is also with the Peng Cheng Laboratory, Shenzhen, China 518066.}
\thanks{L.~Huang is with the College of Computer Science and Technology, Zhejiang University of Technology, Hangzhou, China (email: lianghuang@zjut.edu.cn)}
\thanks{H.~Wang is with the Shenzhen Institute of Information Technology, Shenzhen, China 518172 (email: wanghui@sziit.edu.cn)}
\thanks{Y-J.~A.~Zhang is with the Department of Information Engineering, The Chinese University of Hong Kong, Shatin, N.T., Hong Kong. (e-mail: yjzhang@ie.cuhk.edu.hk).}
\thanks{\textcolor{red}{The complete source code implementing LyDROO is available on-line at https://github.com/revenol/LyDROO.}}}
\maketitle

\begin{abstract}
Opportunistic computation offloading is an effective method to improve the computation performance of mobile-edge computing (MEC) networks under dynamic edge environment. In this paper, we consider a multi-user MEC network with time-varying wireless channels and stochastic user task data arrivals in sequential time frames. In particular, we aim to design an online computation offloading algorithm to maximize the network data processing capability subject to the long-term data queue stability and average power constraints. The online algorithm is practical in the sense that the decisions for each time frame are made without the assumption of knowing the future realizations of random channel conditions and data arrivals. We formulate the problem as a multi-stage stochastic mixed integer non-linear programming (MINLP) problem that jointly determines the binary offloading (each user computes the task either locally or at the edge server) and system resource allocation decisions in sequential time frames. To address the coupling in the decisions of different time frames, we propose a novel framework, named LyDROO, that combines the advantages of Lyapunov optimization and deep reinforcement learning (DRL). Specifically, LyDROO first applies Lyapunov optimization to decouple the multi-stage stochastic MINLP into deterministic per-frame MINLP subproblems. By doing so, it guarantees to satisfy all the long-term constraints by solving the per-frame subproblems that are much smaller in size. Then, LyDROO integrates model-based optimization and model-free DRL to solve the per-frame MINLP problems with very low computational complexity. Simulation results show that under various network setups, the proposed LyDROO achieves optimal computation performance while stabilizing all queues in the system. Besides, it induces very low computation time that is particularly suitable for real-time implementation in fast fading environments.
\end{abstract}


\section{Introduction}
\subsection{Motivations and Summary of Contributions}
\IEEEPARstart{T}{he} emerging mobile-edge computing (MEC) technology is widely recognized as a key solution to enhance the computation performance of wireless devices (WDs) \cite{2017:Mao}, especially for size-constrained IoT (Internet of Things) devices with low on-device battery and computing capability. With MEC servers deployed at the edge of radio access networks, e.g., cellular base stations, WDs can offload intensive computation tasks to the edge server (ES) in the vicinity to reduce the computation energy and time cost. Compared to the naive scheme that offloads all the tasks for edge execution, \emph{opportunistic computation offloading}, which dynamically assigns tasks to be computed either locally or at the ES, has shown significant performance improvement under time-varying network conditions, such as wireless channel gains\cite{2013:Wu}, harvested energy level\cite{2016:You}, task input-output dependency \cite{2020:Yan}, and edge caching availability \cite{2020:Bi}, etc.

There have been extensive studies on opportunistic computation offloading to optimize the computation performance of multi-user MEC networks\cite{2019:Lee,2020:Yan,2020:Bi,2018:Bi,2017:Dinh}. In general, it involves solving a mixed integer non-linear programming (MINLP) that jointly determines the binary offloading (i.e., either offloading the computation or not) and the communication/computation resource allocation (e.g., task offloading time and local/edge CPU frequencies) decisions. Solving such problems typically requires prohibitively high computational complexity especially in large-size networks. Accordingly, many works have focused on designing reduced-complexity sub-optimal algorithms, such as local-search based heuristics \cite{2018:Bi,2020:Yan}, decomposition-oriented search \cite{2018:Bi}, and convex relaxations of the binary variables \cite{2019:Du,2017:Dinh}, etc. However, aside from performance losses, the above sub-optimal algorithms still require a large number of numerical iterations to produce a satisfying solution. In practice, the MINLP needs to be frequently re-solved once the system parameters, such as wireless link quality, vary. It is therefore too costly to implement the conventional optimization algorithms in a highly dynamic MEC environment.

The recent development of data-driven \emph{deep reinforcement learning} (DRL) provides a promising alternative to tackle the online computation offloading problem. In a nutshell, the DRL framework takes a model-free approach that uses deep neural networks (DNNs) to directly learn the optimal mapping from the ``state" (e.g., time-varying system parameters) to the ``action" (e.g., offloading decisions and resource allocation) to maximize the ``reward" (e.g., data processing rate) via repeated interactions with the environment \cite{2018:Sutton}. It eliminates the complicated computation of MINLP and automatically learns from the past experience on-the-fly without requiring manually labeled training data samples, and thus is particularly advantageous for online implementation. Many studies have applied DRL techniques to design online offloading algorithms in MEC networks \cite{2019:Liu,2019:Min,2018:Li,2019:Chen,2020:Tang,2019:Wei,2020:Du,2020:Xiao,2020:Zhang,2019:Huang}. In particular, our previous work \cite{2019:Huang} proposes a hybrid framework, named DROO (Deep Reinforcement learning-based Online Offloading), to combine the advantages of conventional model-based optimization and model-free DRL methods. DROO implements a DNN to produce binary offloading decisions based on the input environment parameters such as channel conditions. The candidate offloading solutions are then fed into a model-based optimization module, which accordingly optimizes the communication/computation resource allocation and outputs an accurate estimate of the reward value for each candidate offloading decision. The integrated learning and optimization approach leads to more robust and faster convergence of the online training process, thanks to the accurate estimation of reward values corresponding to each sampled action.

Apart from optimizing the computation performance, it is equally important to guarantee stable system operation, such as data queue stability and average power consumption. However, most of the existing DRL-based methods do not impose long-term performance constraints (e.g., \cite{2019:Liu,2019:Min,2018:Li,2019:Chen,2020:Tang,2020:Du,2020:Xiao,2019:Wei,2020:Zhang,2019:Huang}). Instead, they resort to heuristic approaches that discourage unfavorable actions in each time frame by introducing penalty terms related to, for example, packet drop events \cite{2019:Chen,2020:Tang} and energy consumption \cite{2019:Min,2020:Zhang}. A well-known framework for online joint utility maximization and stability control is \emph{Lyapunov optimization} \cite{2010:Neely}. It decouples a multi-stage stochastic optimization to sequential per-stage deterministic subproblems, while providing theoretical guarantee to long-term system stability. Some recent works have applied Lyapunov optimization to design computation offloading strategy in MEC networks (e.g., \cite{2019:Du,2016:Mao,2017:Sun,2017:Mao1,2019:Liu1}). However, it still needs to solve a hard MINLP in each per-stage subproblem to obtain the joint binary offloading and resource allocation decisions. To tackle the intractability, some works have designed reduced-complexity heuristics, such as continuous relaxation in \cite{2019:Du} and decoupling heuristic in \cite{2019:Liu1}. This, however, suffers from the similar performance-complexity tradeoff dilemma as in \cite{2020:Yan,2020:Bi,2018:Bi,2017:Dinh}.

In this paper, we consider a multi-user MEC network in Fig.~\ref{101}, where the computation task data arrive at the WDs' data queues stochastically in sequential time frames. We aim to design an online computation offloading algorithm, in the sense that the decisions for each time frame are made without the assumption of knowing the future realizations of random channel conditions and data arrivals. The objective is to maximize the network data processing capability subject to the long-term data queue stability and average power constraints. To tackle the problem, we propose a Lyapunov-guided Deep Reinforcement learning (DRL)-based Online Offloading (LyDROO) framework that combines the advantages of Lyapunov optimization and DRL. Under fast-varying channel fading and dynamic task arrivals, LyDROO can make online optimal decisions in real time, while guaranteeing the long-term system stability. To the authors' best knowledge, this is the first work that combines Lyapunov optimization and DRL for online computation offloading design in MEC networks. The main contributions of the paper are:

\begin{figure}
\centering
  \begin{center}
    \includegraphics[width=0.45\textwidth]{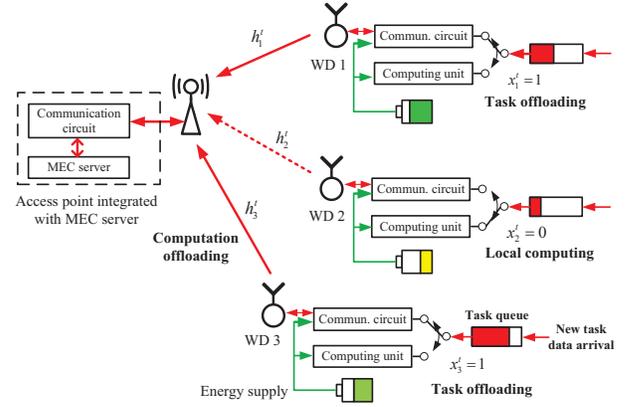}
  \end{center}
  \caption{The considered multi-user MEC network in a tagged time frame.}
  \label{101}
\end{figure}

\begin{itemize}
  \item \emph{Online stable computation offloading design:} Considering random fading channels and data arrivals, we formulate the problem as a multi-stage stochastic MINLP to maximize the long-term average weighted sum computation rate (i.e., the number of processed bits per second) of all the WDs, subject to the queue stability and average power constraints. In particular, we will make the optimal offloading and resource allocation decisions in each time frame without the assumption of knowing the future realizations of random channel conditions and data arrivals.
  \item \emph{Integrated Lyapunov-DRL framework:} To tackle the problem, we propose a novel LyDROO framework that combines the advantages of Lyapunov optimization and DRL. In particular, we first apply Lyapunov optimization to decouple the multi-stage stochastic MINLP into per-frame deterministic MINLP problems. Then in each frame, we integrate model-based optimization and model-free DRL to solve the per-frame MINLP problems with very low computational complexity. In particular, we show that the proposed LyDROO framework not only ensures the long-term queue stability and average power constraints, but also obtains the optimal computation rate performance in an online fashion.
  \item \emph{Integrated optimization and learning:} LyDROO adopts an actor-critic structure to solve the per-frame MINLP problem. The actor module is a DNN that learns the optimal binary offloading action based on the input environment parameters including the channel gains and queue backlogs of all the WDs. The critic module evaluates the binary offloading action by analytically solving the optimal resource allocation problem. Compared to the conventional actor-critic structure that uses a model-free DNN in the critic module, the proposed approach takes advantage of model information to acquire accurate evaluation of the action, and thus enjoying more robust and faster convergence of the DRL training process.
  \item \emph{Balanced exploration and exploitation:} LyDROO deploys a noisy order-preserving quantization method to generate offloading action, which elegantly balances the exploration-exploitation tradeoff (i.e., performance or diversity oriented) in the DRL algorithm design to ensure fast training convergence. Besides, the quantization method can adaptively adjust its parameter during the training process, which yields significant reduction in computational complexity without compromising the convergence performance.
\end{itemize}

Simulation results show that the proposed LyDROO algorithm converges very fast to the optimal computation rate while meeting all the long-term stability constraints. Compared to a myopic benchmark algorithm that greedily maximizes the computation rate in each time frame, the proposed LyDROO achieves a much larger \emph{stable capacity region} that can stabilize the data queues under much heavier task data arrivals and more stringent power constraint.

\subsection{Related Works}
Binary and partial computation offloading are two common offloading models in edge computing systems. While the former requires the entire dataset of a computation task to be processed as a whole either locally at a wireless device (WD) or remotely at the edge server, the latter allows the dataset to be partitioned and executed in parallel at both the WD and the edge server \cite{2017:Mao}. In this paper, we focus on the design of online binary offloading strategy, which is widely adopted in IoT networks for executing simple computation tasks with non-partitionable dataset. Meanwhile, we discuss in Section VII the application of the proposed LyDROO scheme to design online partial offloading strategy when the computation task consists of multiple independent subtasks.

Reduced-complexity algorithms have been widely explored in the literature to tackle the intractability of combinatorial computation offloading problem in multi-user MEC networks adopting binary offloading model. For instance, \cite{2019:Lee} considers WDs offloading their tasks to the neighboring nodes that arrive and departure in random. It formulates an online stopping problem and proposes a low-complexity algorithm, where each WD individually selects the best set of neighboring nodes in an online manner to minimize the worst-case computation latency. The proposed method, however, is not suitable for optimizing a long-term average objective considered in this paper. \cite{2018:Bi} proposes a coordinate descent method that iteratively finds the local-optimum by flipping the binary offloading decision of one user at a time. \cite{2020:Yan} applies Gibbs sampling to search the decision space in a stochastic manner. To reduce the search dimensions, \cite{2018:Bi} proposes an ADMM (alternating direction method of multipliers) based method that decomposes the original combinatorial optimization into parallel one-dimension sub-problems. Besides the search-based meta-heuristic algorithms, existing work has also applied convex relaxation to handle the binary variables, such as linear relaxation \cite{2018:Bi,2019:Du} and quadratic approximation \cite{2017:Dinh}. The aforementioned optimization methods, however, inevitably encounter the performance-complexity tradeoff dilemma when handling integer variables, and are not suitable for online implementation that requires consistently high solution quality under fast-varying environment.

DRL has recently appeared as a promising alternative to solve online computation offloading problems in MEC networks. Existing DRL-based methods take either value-based or policy-based approach to learn the optimal mapping from the ``state" (e.g., time-varying system parameters) to the ``action" (e.g., offloading decisions and resource allocation). Commonly used value-based DRL methods include deep Q-learning network (DQN) \cite{2019:Liu,2019:Min,2018:Li}, double DQN \cite{2019:Chen} and dueling DQN \cite{2020:Tang}, where a DNN is trained to estimate the state-action value function. However, DQN-based methods are costly when the number of possible discrete offloading actions is large, e.g., exponential in the number of WDs. To resolve this issue, recent works have applied policy-based approach, such as the actor-critic DRL \cite{2019:Wei,2019:Huang,2020:Du} and the deep deterministic policy gradient (DDPG) methods \cite{2020:Zhang,2020:Xiao}, to directly construct the optimal mapping policy from the input state to the output action using a DNN. For example, \cite{2020:Xiao} considers a WD taking only discrete offloading actions, including integer offloading decision and discredited transmit power and offloading rate, and applies an actor-critic DRL method to learn the optimal mapping from continuous input state to the discrete output actions. \cite{2019:Huang} and \cite{2020:Zhang} train two separate learning modules to generate discrete offloading decision and continuous resource allocation sequentially. Specifically, \cite{2020:Zhang} applies an actor DNN to generate the resource allocation solution, concatenated by a DQN-based critic network to select the discrete offloading action. Similar to \cite{2019:Liu,2019:Min,2018:Li,2019:Chen,2020:Tang}, the estimation of state-action value function in the critic network is difficult when the number of possible offloading actions is large. On the other hand, the DROO framework proposed in \cite{2019:Huang} uses an actor DNN to generate a small number of binary offloading decisions, followed by a model-based critic module that selects the best action by analytically solving the optimal resource allocation problem. Thanks to the accurate evaluation of action acquired by the critic module, DROO enjoys fast convergence to the optimal solution even when the actor DNN provides very few actions (e.g., two actions after sufficient iterations) for the critic to select from. In this paper, we embed DROO in the LyDROO framework to solve the per-frame MINLP problems.

The above DRL-based methods fail to address the long-term performance requirements, e.g., queue stability and average power, under random environments. In this regard, recent studies have applied Lyapunov optimization to design an online offloading strategy with long-term performance guarantee \cite{2019:Du,2016:Mao,2017:Sun,2017:Mao1,2019:Liu1}. Lyapunov optimization decouples the multi-stage stochastic problem to per-frame deterministic subproblems. For each per-frame subproblem, \cite{2016:Mao} considers the binary offloading decision of a single WD. Likewise, \cite{2017:Sun} schedules only one user to offload to one of the multiple ESs in each time frame. In both cases, the number of binary offloading variables is very small, and hence the optimal solution can be obtained by brute force search. \cite{2017:Mao1,2019:Du,2019:Liu1} consider joint offloading decisions of multiple users. Unlike the binary offloading policy considered in this paper, \cite{2017:Mao1} allows the WDs to process task data in parallel both locally and at the ES, and applies convex optimization to solve the continuous joint offloading and resource allocation problem. In contrast, \cite{2019:Du} and \cite{2019:Liu1} adopt binary offloading policy where the number of possible offloading solutions grows exponentially with the user number. To tackle the combinatorial problem, \cite{2019:Du} relaxes the binary variables into continuous ones. \cite{2019:Liu1} proposes a two-stage heuristic, which first fixes the resource allocation and then obtains the binary offloading decisions using matching theory. However, these heuristic methods cannot guarantee consistently high solution quality, which may eventually degrade the long-term performance.

\begin{figure}
\centering
  \begin{center}
    \includegraphics[width=0.45\textwidth]{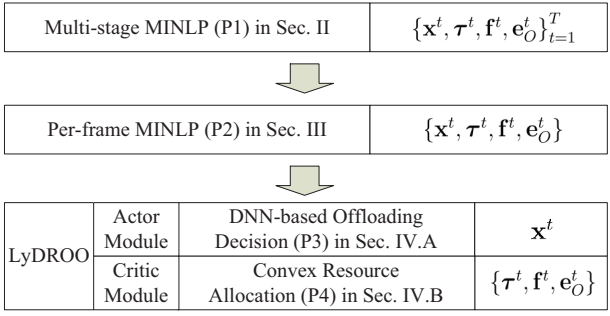}
  \end{center}
  \caption{Organization of the paper.}
  \label{102}
\end{figure}

In Fig.~\ref{102}, we illustrate the organization of the rest of the paper. In Section II, we formulate the stable computation offloading problem as a multi-stage stochastic MINLP problem (P1). In Section III, we apply the Lyapunov optimization to decouple (P1) into per-frame deterministic MINLP subproblem (P2). In Section IV, we introduce the LyDROO algorithm to solve (P2) using an actor-critic DRL. The actor module implements a DNN to solve the binary offloading subproblem (P3) and the critic module applies a customized optimization algorithm to solve the continuous resource allocation problem (P4). In Section V, we analyze the performance of the LyDROO algorithm. In Section VI, we evaluate the proposed algorithm via extensive simulations. Finally, we conclude the paper in Section VII.

\section{System Model and Problem Formulation}
\subsection{System Model}
As shown in Fig.~\ref{101}, we consider an ES assisting the computation of $N$ WDs in sequential time frames of equal duration $T$. Within the $t$th time frame, we denote $A_{i}^t$ (in bits) as the raw task data arrival at the data queue of the $i$th WD. We assume that the arrival $A_{i}^t$ follows a general i.i.d. distribution with bounded second order moment, i.e., $\mathbb{E}\left[\left(A_{i}^t\right)^2\right] = \eta_i < \infty$, for $i=1,\cdots,N$. We assume that the value of $\eta_i$ is known, e.g., by estimating from past observations. We denote the channel gain between the $i$th WD and the ES as $h^t_i$. Under the block fading assumption, $h^t_i$ remains constant within a time frame but varies independently across different frames.

In the $t$th time frame, suppose that a tagged WD $i$ processes $D_{i}^t$ bits data and produces a computation output at the end of the time frame. In particular, we assume that the WDs adopt a binary computation offloading rule \cite{2017:Mao}. That is, within each time frame, the raw data must be processed either locally at the WD or remotely at the ES. For instance, WD $1$ and $3$ offload their tasks while WD $2$ computes locally in Fig.~\ref{101}. The offloading WDs share a common bandwidth $W$ for transmitting the task data to the ES in a TDMA manner. We use a binary variable $x^t_i$ to denote the offloading decision, where $x^t_i=1$ and $0$ denote that WD $i$ performs computation offloading and local computing, respectively.

When the WD processes the data locally ($x^t_i=0$), we denote the local CPU frequency as $f^t_i$, which is upper bounded by $f_i^{max}$. The raw data (in bits) processed locally and the consumed energy within the time frame are \cite{2017:Mao}
\begin{equation}
\label{51}
D^{t}_{i,L} = f^t_i T/\phi,\ E^{t}_{i,L} = \kappa \left(f^t_i\right)^3 T, \ \forall x_{i}^t =0,
\end{equation}
respectively. Here, parameter $\phi>0$ denotes the number of computation cycles needed to process one bit of raw data and $\kappa>0$ denotes the computing energy efficiency parameter.

Otherwise, when the data is offloaded for edge execution ($x^t_i=1$), we denote $P_i^t$ as the transmit power constrained by the maximum power $P_i^t \leq P^{max}_i$ and $\tau^t_i T$ as the amount of time allocated to the $i$th WD for computation offloading. Here, $\tau^t_i \in [0,1]$ and $\sum_{i=1}^N \tau^t_i \leq 1$. The energy consumed on data offloading is $E^{t}_{i,O} = P_i^t \tau^t_i T$. Similar to \cite{2016:You} and \cite{2018:Bi}, we neglect the delay on edge computing and result downloading such that the amount of data processed at the edge within the time frame is
\begin{equation}
\label{52}
\begin{aligned}
D^{t}_{i,O} &=  \frac{W\tau^t_i T}{v_u}\log_2\left(1+\frac{P_i^t h_i^t}{N_0}\right) \\
& =   \frac{W\tau^t_i T}{v_u}\log_2\left(1+\frac{E_{i,O}^t h_i^t}{\tau^t_i T N_0}\right), \ \forall x_{i}^t =1,
\end{aligned}
\end{equation}
where $v_u\geq 1$ denotes the communication overhead and $N_0$ denotes the noise power.

Let $D^{t}_{i} \triangleq (1 - x^t_i) D^{t}_{i,L} + x^t_i D^{t}_{i,O}$ and $E^t_i \triangleq (1 - x^t_i) E^{t}_{i,L} + x^t_i E^{t}_{i,O}$ denote the bits computed and energy consumed in time frame $t$. We define \emph{computation rate} $r_i^t$ and \emph{power consumption} $e_i^t$ in the $t$th time frame as
\begin{equation}
\label{14}
\begin{aligned}
r_i^t &= \frac{D^{t}_{i}}{T} =\frac{(1 - x^t_i)f^t_i}{\phi} + x^t_i \frac{W\tau^t_i}{v_u}\log_2\left(1+\frac{e^t_{i,O} h_i^t}{\tau^t_i N_0}\right), \\
e_i^t &= \frac{E^{t}_{i}}{T} = (1 - x^t_i) \kappa \left(f^t_i\right)^3 + x^t_i e^t_{i,O},
\end{aligned}
\end{equation}
where $e^t_{i,O} \triangleq  E^{t}_{i,O}/T $. For simplicity of exposition, we assume $T=1$ without loss of generality in the following derivations.

Let $Q_{i}(t)$ denote the queue length of the $i$th WD at the beginning of the $t$th time frame. Then, the queue dynamics can be modeled as
\begin{equation}
Q_{i}(t+1) = \max\left\{Q_{i}(t) - \tilde{D}^{t}_{i} + A_{i}^t ,0\right\}, \ i=1,2,\cdots,
\end{equation}
where $\tilde{D}^{t}_{i} = \min \left(Q_{i}(t),D^{t}_{i}\right)$ and $Q_{i}(1)=0$. In this paper, we consider infinite queueing capacity for analytical tractability. In the following derivation, we enforce the data causality constraint $D^{t}_{i} \leq Q_{i}(t)$, implying that $Q_i(t)\geq 0$ holds for any $t$. Thus, the queue dynamics is simplified as
\begin{equation}
\label{111}
Q_i(t+1) = Q_{i}(t) - D^{t}_{i} + A_{i}^t,\ \ i=1,2,\cdots.
\end{equation}

\textbf{Definition 1}: A discrete time queue $Q_{i}(t)$ is \emph{strongly stable} if the time average queue length $\lim_{K\rightarrow \infty} \frac{1}{K} \mathsmaller\sum_{t=1}^K \mathbb{E} \left[Q_i(t)\right] <\infty$, where the expectation is taken with respect to the system random events \cite{2010:Neely}, i.e., channel fading and task data arrivals in this paper.

By the Little's law, the average delay is proportional to the average queue length. Thus, a strongly stable data queue translates to a finite processing delay of each task data bit.

\subsection{Problem Formulation}
In this paper, we aim to design an online algorithm to maximize the long-term average weighted sum computation rate of all the WDs under the data queue stability and average power constraints. In particular, we make online decisions in the sense that in each time frame, we optimize the task offloading and the resource allocation decisions for the particular time frame without the assumption of knowing the future realizations of random channel conditions and data arrivals. We denote $\mathbf{x}^t = \left[x_1^t,\cdots,x_N^t\right]$, $\boldsymbol{\tau}^t = \left[\tau_1^t,\cdots,\tau_N^t\right]$, $\mathbf{f}^t = \left[f_1^t,\cdots,f_N^t\right]$ and $\mathbf{e}_O^t = \left[e_{1,O}^t,\cdots,e_{N,O}^t\right]$, and let $\mathbf{x} = \left\{\mathbf{x}^t\right\}_{t=1}^K$, $\boldsymbol{\tau} = \left\{\boldsymbol{\tau}^t\right\}_{t=1}^K$, $\mathbf{f}= \left\{\mathbf{f}^t\right\}_{t=1}^K$ and $\mathbf{e}_O = \left\{\mathbf{e}_O^t \right\}_{t=1}^K$. We formulate the problem as the following multi-stage stochastic MINLP problem (P1):
\begin{subequations}
   \label{6}
   \begin{align}
    &\underset{\mathbf{x}, \boldsymbol{\tau},\mathbf{f},\mathbf{e}_O}{\text{maximize}}  \lim_{K\rightarrow \infty} \frac{1}{K} \cdot \mathsmaller\sum_{t=1}^K \mathsmaller\sum_{i=1}^N c_i r_i^t \notag\\
    & \text{subject to} \notag \\
    &\mathsmaller\sum_{i=1}^N \tau^t_i  \leq 1, \ \forall t, \label{62}\\
    & (1-x^t_i)f^t_i /\phi + x^t_i\frac{W\tau^t_i}{v_u}\log_2\left(1+\frac{e^t_{i,O} h_i^t}{\tau^t_i N_0}\right) \leq Q_i(t) , \ \forall i, t, \label{65}\\
    & \lim_{K\rightarrow \infty} \frac{1}{K} \cdot \mathsmaller\sum_{t=1}^K \mathbb{E}\left[(1 - x^t_i) \kappa \left(f^t_i\right)^3 + x^t_i e^t_{i,O}\right] \leq \gamma_i,\  \forall i,\label{63}\\
    & \lim_{K\rightarrow \infty} \frac{1}{K}\cdot \mathsmaller\sum_{t=1}^K \mathbb{E} \left[Q_i(t)\right] <\infty, \forall i, \label{64}\\
    & f^t_i \leq f^{max}_i,\ \ e^t_{i,O}\leq P^{max}_i \tau^t_i,\ \forall i, t,\\
        & x^t_i \in \left\{0,1\right\},\  \tau^t_i, f^t_i,e^t_{i,O} \geq 0,\ \forall i, t.
   \end{align}
\end{subequations}
Here, $c_i$ denotes the fixed weight of the $i$th WD. (\ref{62}) denotes the offloading time constraint. Notice that $\tau^t_i = e^t_{i,O}=0$ must hold at the optimum if $x^t_i=0$. Similarly, $f^t_i=0$ must hold if $x^t_i=1$. (\ref{65}) corresponds to the data causality constraint. (\ref{63}) corresponds to the average power constraint and $\gamma_i$ is the power threshold. (\ref{64}) are the data queue stability constraints. Under the stochastic channels and data arrivals, it is hard to satisfy the long-term constraints when the decisions are made in each time frame without knowing the future realizations of random channel conditions and data arrivals. Besides, the fast-varying channel condition requires real-time decision-making in each short time frame, e.g., within the channel coherence time. In the following, we propose a novel LyDROO framework that solves (P1) with both high robustness and efficiency.

\emph{Remark 1:} Before leaving this session, we comment on the possible extension of the proposed LyDROO algorithm. (P1) uses a linear utility function $U\left(r^t_i\right) = r^t_i$ in the objective. However, we will show later in Section IV that the proposed LyDROO framework is applicable to solve a wide range of problems as long as the resource allocation problem (P4) can be efficiently solved. For instance, we can consider a general non-decreasing concave function $U\left(r^t_i\right)$ such that the corresponding (P4) is a convex problem, e.g., $\alpha$-fairness function $(1-\alpha)^{-1}\left(r^t_i\right)^{1-\alpha}$ with $\alpha\geq 0$ and $\alpha\neq 1$, proportional fairness function $\ln(r^t_i)$, or other suitable QoS (quality of service) utilities (see \cite{2017:Yang} and the reference therein). For analytical clarity, we consider in this paper a specific linear utility function to highlight the features of the LyDROO framework.

\section{Lyapunov-based Decoupling of the Multi-stage MINLP}
In this section, we apply the Lyapunov optimization to decouple (P1) into per-frame deterministic problems. To cope with the average power constraints (\ref{63}), we introduce $N$ virtual energy queues $\left\{Y_i(t)\right\}_{i=1}^N$, one for each WD. Specifically, we set $Y_i(1) =0$ and update the queue as
\begin{equation}
\label{112}
Y_i(t+1) = \max \left(Y_i(t) + \nu e^t_i - \nu \gamma_i, 0 \right),
\end{equation}
for $i=1,\cdots,N$ and $t=1,\cdots,K$, where $e^t_i$ in (\ref{14}) is the energy consumption at the $t$th time frame and $\nu$ is a positive scaling factor. $Y_i(t)$ can be viewed as a queue with random ``energy arrivals" $\nu e^t_i$ and fixed ``service rate" $\nu \gamma_i$. Intuitively, when the virtual energy queues are stable, the average power consumption $e_i^t$ (i.e., the virtual queue arrival rate) does not exceed $\gamma_i$, and thus the constraints in (\ref{63}) are satisfied.

To jointly control the data and energy queues, we define $\mathbf{Z}(t) = \left\{\mathbf{Q}(t),\mathbf{Y}(t)\right\}$ as the total queue backlog, where $\mathbf{Q}(t) = \left\{Q_i(t)\right\}_{i=1}^N$ and $\mathbf{Y}(t) = \left\{Y_i(t)\right\}_{i=1}^N$. Then, we introduce the Lyapunov function $L\left( \mathbf{Z}(t)\right)$ and Lyapunov drift $\Delta L\left( \mathbf{Z}(t)\right)$ as  \cite{2010:Neely}
\begin{equation}
\begin{aligned}
L\left( \mathbf{Z}(t)\right) &= 0.5\left( \mathsmaller\sum_{i=1}^N Q_i(t)^2 +  \mathsmaller\sum_{i=1}^N Y_i(t)^2\right), \\ \Delta L\left( \mathbf{Z}(t)\right) &= \mathbb{E}\left\{L\left(\mathbf{Z}(t+1)\right)-L\left(\mathbf{Z}(t)\right)| \mathbf{Z}(t) \right\}.
\end{aligned}
\end{equation}
To maximize the time average computation rate while stabilizing the queue $\mathbf{Z}(t)$, we use the drift-plus-penalty minimization approach \cite{2006:Georgiadis}. Specifically, we seek to minimize an upper bound on the following drift-plus-penalty expression at every time frame $t$:
\begin{equation}
\label{57}
\Lambda\left(\mathbf{Z}(t)\right) \triangleq \Delta L\left( \mathbf{Z}(t)\right) - V \cdot \mathsmaller\sum_{i=1}^N \mathbb{E} \left\{ c_i r_i^t|\mathbf{Z}(t)\right\},
\end{equation}
where $V >0$ is an ``importance" weight to scale the penalty.

In the following, we derive an upper bound of $\Lambda\left(\mathbf{Z}(t)\right)$. To begin with, we have
\begin{equation*}
\label{11}
\begin{aligned}
Q_i(t+1)^2 & =   Q_{i}(t)^2 + 2Q_{i}(t)\left(A_{i}^t-D^{t}_{i}\right) + \left(A_{i}^t-D^{t}_{i}\right)^2,\\
Y_i(t+1)^2 & =   Y_i(t)^2  + 2  Y_i(t)\left(e^t_i -\gamma_i\right) + \left( e^t_i - \gamma_i\right)^2.
\end{aligned}
\end{equation*}
By taking the sum over the $N$ queues on both sides, we have
\begin{equation}
\begin{aligned}
&0.5 \mathsmaller\sum_{i=1}^N Q_i(t+1)^2   - 0.5 \mathsmaller\sum_{i=1}^N Q_{i}(t)^2 \\
 =& 0.5 \mathsmaller\sum_{i=1}^N  \left(A_{i}^t-D^{t}_{i}\right)^2  + \mathsmaller\sum_{i=1}^N Q_{i}(t)\left(A_{i}^t-D^{t}_{i}\right) \label{53}
\end{aligned}
\end{equation}
and
\begin{equation}
\begin{aligned}
&0.5\mathsmaller\sum_{i=1}^N  Y_i(t+1)^2  - 0.5 \mathsmaller\sum_{i=1}^N  Y_i(t)^2 \\
 = & 0.5 \mathsmaller\sum_{i=1}^N \left( e^t_i-\gamma_i\right)^2  + \mathsmaller\sum_{i=1}^N  Y_i(t)\left(e^t_i -\gamma_i\right). \label{54}
\end{aligned}
\end{equation}

We define
\begin{equation}
L\left(\mathbf{Q}(t)\right) \triangleq 0.5 \mathsmaller\sum_{i=1}^N Q_i(t)^2
\end{equation}
and
\begin{equation}
\Delta L\left(\mathbf{Q}(t)\right) \triangleq  \mathbb{E}\left\{L\left(\mathbf{Q}(t+1)\right)-L\left(\mathbf{Q}(t)\right)| \mathbf{Z}(t)\right\}.
\end{equation}
By taking the conditional expectation on both sides of (\ref{53}), we have
\begin{equation}
\label{55}
\begin{aligned}
\Delta L\left(\mathbf{Q}(t)\right) \leq B_1 + \mathsmaller\sum_{i=1}^N  Q_{i}(t) \mathbb{E}\left[\left(A_{i}^t-D^{t}_{i}\right)| \mathbf{Z}(t)\right].
\end{aligned}
\end{equation}
Here, $B_1$ is a constant obtained as
\begin{subequations}
\allowdisplaybreaks
\begin{align*}
&0.5 \mathsmaller\sum_{i=1}^N \mathbb{E}\left[\left(A_{i}^t-D^{t}_{i}\right)^2\right]  \leq 0.5 \mathsmaller\sum_{i=1}^N  \mathbb{E}\left[ \left(A_{i}^t\right)^2 + \left(D_{i}^t\right)^2 \right]\\
 & \leq 0.5 \mathsmaller\sum_{i=1}^N \left(\eta_i + \left[T\max\left\{f^{max}_i/\phi, r^{max}_i\right\}\right]^2\right) \triangleq B_1,
\end{align*}
\end{subequations}
where the second inequality holds because $r^{max}_i \triangleq \mathbb{E}\left[\frac{W}{v_u}\log_2\left(1+\frac{P_i^{max} h_i^t}{ N_0}\right)\right]$ corresponds to the maximum average transmission rate of the $i$th WD.

Similarly, we define
\begin{equation}
L\left( \mathbf{Y}(t)\right) = 0.5 \mathsmaller\sum_{i=1}^N Y_i(t)^2
\end{equation}
and
\begin{equation}
\Delta L\left(\mathbf{Y}(t)\right) \triangleq  \mathbb{E}\left\{L\left(\mathbf{Y}(t+1)\right)-L\left(\mathbf{Y}(t)\right)| \mathbf{Z}(t)\right\}.
\end{equation}
We obtain the following by taking the expectation on both sides of (\ref{54})
\begin{equation}
\label{56}
\begin{aligned}
\Delta L\left(\mathbf{Y}(t)\right) & \leq  B_2 + \mathsmaller\sum_{i=1}^N  Y_i(t) \mathbb{E}\left[e^t_i -\gamma_i| \mathbf{Z}(t)\right],
\end{aligned}
\end{equation}
where the constant $B_2$ is obtained from
\begin{equation*}
\begin{aligned}
&0.5 \mathsmaller\sum_{i=1}^N \mathbb{E}\left[\left( e^t_i-\gamma_i\right)^2\right] \\
&\leq 0.5 \mathsmaller\sum_{i=1}^N \left[\left(\max\left\{\kappa\left( f^{max}_i\right)^3,P^{max}_i\right\}\right)^2 + \gamma_i^2\right] \triangleq B_2.
\end{aligned}
\end{equation*}
Summing over the two inequalities in (\ref{55}) and (\ref{56}), we have
\begin{equation}
\begin{aligned}
\Delta L\left(\mathbf{Z}(t)\right)  \leq &\hat{B}  + \mathsmaller\sum_{i=1}^N  Q_{i}(t) \mathbb{E}\left[\left(A_{i}^t-D^{t}_{i}\right)| \mathbf{Z}(t)\right] \\
& + \mathsmaller\sum_{i=1}^N  Y_i(t) \mathbb{E}\left[e^t_i -\gamma_i| \mathbf{Z}(t)\right]
\end{aligned}
\end{equation}
where $\hat{B} =B_1 + B_2$. Therefore, the upper bound of the drift-plus-penalty expression in (\ref{57}) is
\begin{equation}
\label{12}
\begin{aligned}
\hat{B}  &+ \mathsmaller\sum_{i=1}^N  \big\{Q_{i}(t) \mathbb{E}\left[\left(A_{i}^t-D^{t}_{i}\right)| \mathbf{Z}(t)\right] \\
&\ \ \ \ \ \ \ \ \ +  Y_i(t) \mathbb{E}\left[e^t_i -\gamma_i| \mathbf{Z}(t)\right] - V \mathbb{E} \left[ c_i r_i^t|\mathbf{Z}(t)\right] \big\}.
\end{aligned}
\end{equation}

In the $t$th time frame, we apply the technique of \emph{opportunistic expectation minimization}\cite{2010:Neely}. That is, we observe the queue backlogs $\mathbf{Z}(t)$ and decide the joint offloading and resource allocation control action accordingly to minimize the upper bound in (\ref{12}). Notice that only the second term is related to the control action in the $t$th time frame. By removing the constant terms from the observation at the beginning of the $t$th time frame, the algorithm decides the actions by maximizing the following:
\begin{equation}
\mathsmaller\sum_{i=1}^N \left(Q_{i}(t) + V c_i\right) r^{t}_{i} - \mathsmaller\sum_{i=1}^N Y_i(t)e^t_i,
\end{equation}
where $r^{t}_{i}$ and $e^t_i$ are in (\ref{14}). Intuitively, it tends to increase the computation rates of WDs that have a long data queue backlog or a large weight, while penalizing those that have exceeded the average power threshold.  We introduce an auxiliary variable $r_{i,O}^t$ for each WD $i$ and denote $\mathbf{r}_{O}^t = \left\{r_{i,O}^t\right\}_{i=1}^N$. Taking into account the per-frame constraints, we solve the following deterministic per-frame subproblem (P2) in the $t$th time frame
\begin{subequations}
   \label{7}
   \begin{align}
    &\underset{\mathbf{x}^t, \boldsymbol{\tau}^t,\mathbf{f}^t,\mathbf{e}_O^t, \mathbf{r}_{O}^t}{\text{maximize}} & & \mathsmaller\sum_{i=1}^N \left(Q_{i}(t) + V c_i\right) r^{t}_{i} - \mathsmaller\sum_{i=1}^N Y_i(t)e^t_i \notag \\
    & \text{subject to} & & \mathsmaller \sum_{i=1}^N \tau^t_i  \leq 1, \label{25}\\
    & & & f^t_i /\phi \leq  Q_i(t), \ r_{i,O}^t \leq  Q_i(t),\ \forall i, \label{78}\\
    & & & r_{i,O}^t \leq \frac{ W\tau^t_i}{v_u}\log_2\left(1+\frac{e^t_{i,O} h_i^t}{\tau^t_i N_0}\right), \ \forall i, \label{79}\\
    & & & f^t_i \leq f^{max}_i,\ e^t_{i,O}\leq P^{max}_i \tau^t_i, \ \forall i, \\
    & & & x^t_i \in \left\{0,1\right\},\ \tau^t_i, f^t_i,e^t_{i,O} \geq 0, \ \forall i. \label{85}
   \end{align}
\end{subequations}
Notice that the above constraints (\ref{78}) and (\ref{79}) are equivalent to (\ref{65}) in (P1), because there is exactly one non-zero term in the left-hand side of (\ref{65}) at the optimum. In Section V, we will show that we can satisfy all long-term constraints in (P1) by solving the per-frame subproblems in an online fashion. Then, the remaining difficulty lies in solving the MINLP (P2) in each time frame. In the following section, we propose a DRL-based algorithm to solve (P2) efficiently.

\section{Lyapunov-guided DRL for Online Computation Offloading}
Recall that to solve (P2) in the $t$th time frame, we observe $\boldsymbol{\xi}^t \triangleq \left\{h^t_i, Q_i(t), Y_i(t)\right\}_{i=1}^N$, consisting of the channel gains $\left\{h^t_i\right\}_{i=1}^N$ and the system queue states $\left\{Q_i(t), Y_i(t)\right\}_{i=1}^N$, and accordingly decide the control action $\left\{\mathbf{x}^t,\mathbf{y}^t\right\}$, including the binary offloading decision $\mathbf{x}^t$ and the continuous resource allocation $\mathbf{y}^t \triangleq \left\{\tau_i^t, f_i^t, e_{i,O}^t, r_{i,O}^t\right\}_{i=1}^N$. A close observation shows that although (P2) is a non-convex optimization problem, the resource allocation problem to optimize $\mathbf{y}^t$ is in fact an ``easy" convex problem if $\mathbf{x}^t$ is fixed. In Section IV.B, we will propose a customized algorithm to efficiently obtain the optimal $\mathbf{y}^t$ given $\mathbf{x}^t$ in (P2). Here, we denote $G\left(\mathbf{x}^t,\boldsymbol{\boldsymbol{\xi}}^t\right)$ as the optimal value of (P2) by optimizing $\mathbf{y}^t$ given the offloading decision $\mathbf{x}^t$ and parameter $\boldsymbol{\boldsymbol{\xi}}^t$. Therefore, solving (P2) is equivalent to finding the optimal offloading decision $\left(\mathbf{x}^t\right)^*$, where
\begin{equation}
\label{94}
    (\textrm{P3}):\ \  \left(\mathbf{x}^t\right)^* =  \arg \underset{\mathbf{x}^t \in \{0,1\}^N}{\text{maximize}} \ \ \ G\left(\mathbf{x}^t,\boldsymbol{\boldsymbol{\xi}}^t\right).
\end{equation}

\begin{figure*}
\centering
\includegraphics[width=0.85 \textwidth]{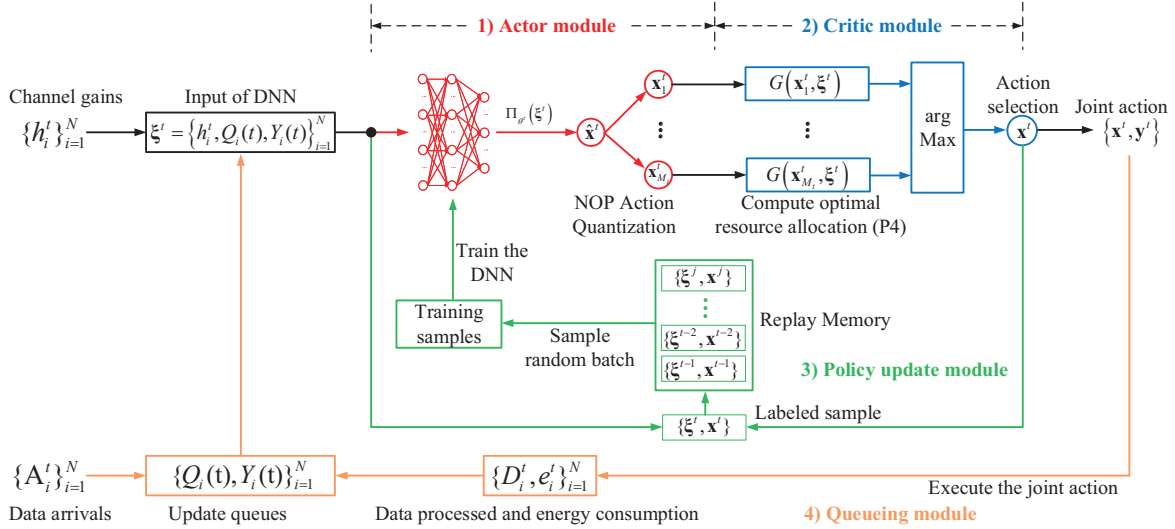}
\caption{The schematics of the proposed LyDROO algorithm.}
\label{fig:network}
\end{figure*}

In general, obtaining $\left(\mathbf{x}^t\right)^*$ requires enumerating $2^N$ offloading decisions, which leads to significantly high computational complexity even when $N$ is moderate (e.g., $N=10$). Other search based methods, such as branch-and-bound and block coordinate descent \cite{1998:Papadimitriou}, are also time-consuming when $N$ is large. In practice, neither method is applicable to online decision-making under fast-varying channel condition. Leveraging the DRL technique, we propose a LyDROO algorithm to construct a policy $\pi$ that maps from the input $\boldsymbol{\boldsymbol{\xi}}^t$ to the optimal action $\left(\mathbf{x}^t\right)^*$, i.e., $\pi : \boldsymbol{\boldsymbol{\xi}}^t \mapsto \left(\mathbf{x}^t\right)^*$, with very low complexity, e.g., tens of milliseconds computation time (i.e., the time duration from observing $\boldsymbol{\xi}^t$ to producing a control action $\left\{\mathbf{x}^t,\mathbf{y}^t\right\}$) when $N=10$.

\subsection{Algorithm Description}
As illustrated in Fig.~\ref{fig:network}, LyDROO consists of four main modules: an actor module that accepts the input $\boldsymbol{\boldsymbol{\xi}}^t$ and outputs a set of candidate offloading actions $\left\{\mathbf{x}^t_i\right\}$, a critic module evaluates $\left\{\mathbf{x}^t_i\right\}$ and selects the best offloading action $\mathbf{x}^t$, a policy update module improves the policy of the actor module over time, and a queueing module updates the system queue states $\left\{Q_i(t), Y_i(t)\right\}_{i=1}^N$ after executing the offloading actions. Through repeated interactions with the random environment $\left\{h^t_i,A^t_i\right\}_{i=1}^N$, the four modules operate in a sequential and iterative manner as detailed below.

\subsubsection{Actor Module} The actor module consists of a DNN and an action quantizer. At the beginning of the $t$th time frame, we denote the parameter of the DNN as $\boldsymbol{\theta}^t$, which is randomly initialized following the standard normal distribution when $t=1$. Taking the observation $\boldsymbol{\boldsymbol{\xi}}^t$ as the input, the DNN outputs a relaxed offloading decision $\mathbf{\hat{x}}^t \in [0,1]^{N}$ that is later to be quantized into feasible binary actions. The input-output relation is expressed as
\begin{equation}
\label{DNN}
\Pi_{\boldsymbol{\theta}^t}: \boldsymbol{\boldsymbol{\xi}}^t \mapsto \mathbf{\hat{x}}^t = \left\{\hat{x}^t_i \in [0,1], i=1,\cdots,N \right\}.
\end{equation}
The well-known universal approximation theorem claims that a multi-layer perceptron with a sufficient number of neurons can accurately approximate any continuous mappings if proper activation functions are applied at the neurons, e.g., sigmoid, ReLu, and tanh functions \cite{marsland2015ml}. Here, we use a sigmoid activation function at the output layer.

We then quantize the continuous $\mathbf{\hat{x}}^t$ into $M_t$ feasible candidate binary offloading actions, where $M_t$ is a time-dependent design parameter. The quantization function is expressed as:
\begin{equation}
\label{quantize}
\Upsilon_{M_t}: \mathbf{\hat{x}}^t \mapsto  \Omega^t = \left\{\mathbf{x}^t_j | \mathbf{x}^t_j \in \{0,1\}^{N}, j=1,\cdots,M_t \right\},
\end{equation}
where $\Omega_t$ denotes the set of candidate offloading actions in the $t$th time frames. $\Upsilon_{M_t}$ represents a quantization function that generates $M_t = |\Omega_t|$ binary actions. A good quantization function should balance the \emph{exploration-exploitation tradeoff} in generating the offloading action to ensure good training convergence. Intuitively, $\left\{\mathbf{x}^t_j\right\}$'s should be close to $\mathbf{\hat{x}}^t$ (measured by Euclidean distance) to make effective use of the DNN's output and meanwhile sufficiently separate to avoid premature convergence to sub-optimal solution in the training process.

Here, we apply the noisy order-preserving (NOP) quantization method \cite{2020:YanDRL}, which can generate any $M_t\leq 2N$ candidate actions. The NOP method generates the first $M_t/2$ actions ($M_t$ is assumed an even number) by applying the order-preserving quantizer (OPQ) in \cite{2019:Huang} to $\hat{\mathbf{x}}^t$. Specifically, the $1$st action $\mathbf{x}^t_1 = [x^t_{1,1},\cdots,x^t_{1,N}]$ is calculated as
\begin{equation}
\label{95}
\begin{aligned}
x^t_{1,i} =  \begin{cases}
                          1 & \hat{x}^t_{i} > 0.5, \\
                          0 & \hat{x}^t_{i} \leq 0.5,
                   \end{cases}
\end{aligned}
\end{equation}
for $i=1, \cdots, N$. To generate the next $M_t/2-1$ actions, we order the entries of $\hat{\mathbf{x}}^t$ based on the distance to $0.5$, such that $\lvert \hat{x}_{(1)}^t-0.5 \rvert \leq \lvert \hat{x}_{(2)}^t-0.5 \rvert \leq \dots \leq \lvert \hat{x}^t_{(i)}-0.5 \rvert \dots \leq \lvert \hat{x}^t_{(N)}-0.5 \rvert$, where $\hat{x}^t_{(i)}$ denotes the $i$th ordered entry of $\hat{\mathbf{x}}^t$. Then, $\hat{x}^t_{(i)}$'s are used as the decision thresholds to quantize $\hat{\mathbf{x}}^t$, where the $m$th action $\mathbf{x}^t_m$, for $m=2,\cdots, M_t/2$, is obtained from entry-wise comparisons of $\hat{\mathbf{x}}^t$ and $\hat{x}^t_{(m-1)}$. That is,
\begin{align}
\allowdisplaybreaks
\label{96}
x^t_{m,i} =  \begin{cases}
                          1, & \hat{x}^t_{i} > \hat{x}^t_{(m-1)}\\
                          & \textrm{or}\ \left\{\hat{x}^t_{i} = \hat{x}^t_{(m-1)}\ \textrm{and}\ \hat{x}^t_{(m-1)} \leq 0.5 \right\},\\
                          0, & \hat{x}^t_{i} < \hat{x}^t_{(m-1)} \\
                          & \textrm{or}\ \left\{\hat{x}^t_{i} = \hat{x}^t_{(m-1)}\ \textrm{and}\ \hat{x}^t_{(m-1)} > 0.5 \right\},
                   \end{cases}
\end{align}
for $i=1, \cdots, N$. To obtain the remaining $M_t/2$ actions, we first generate a noisy version of $\hat{\mathbf{x}}^t$ denoted as $\tilde{\mathbf{x}}^t = \text{Sigmoid}\left(\hat{\mathbf{x}}^t + \mathbf{n}\right)$, where the random Gaussian noise $\mathbf{n}\sim \mathcal{N}\left(\mathbf{0},\mathbf{I}_{N}\right)$ with $\mathbf{I}_{N}$ being an identity matrix, and $\text{Sigmoid}\left(\cdot\right)$ is the element-wise Sigmoid function that bounds each entry of $\tilde{\mathbf{x}}^t$ within $(0,1)$. Then, we produce the remaining $M_t/2$ actions $\mathbf{x}^t_m$, for $m= M_t/2+1 ,\cdots, M_t$, by applying the OPQ to $\tilde{\mathbf{x}}^t$, i.e., replacing $\hat{\mathbf{x}}^t$ with $\tilde{\mathbf{x}}^t$ in (\ref{95}) and (\ref{96}).

\subsubsection{Critic Module} Followed by the actor module, the critic module evaluates $\left\{\mathbf{x}^t_i\right\}$ and selects the best offloading action $\mathbf{x}^t$. Unlike the conventional actor-critic structure that uses a model-free DNN as the critic network to evaluate the action, LyDROO leverages the model information to evaluate the binary offloading action by analytically solving the optimal resource allocation problem. This enables the critic module to have accurate evaluation of the offloading actions, and thus achieving more robust and faster convergence of the DRL training process.

Specifically, LyDROO selects the best action $\mathbf{x}^t$ as
\begin{equation}
\mathbf{x}^t = \arg \max_{\mathbf{x}^t_j \in \Omega_t} G\left(\mathbf{x}_j^t,\boldsymbol{\boldsymbol{\xi}}^t\right),
\end{equation}
where $G\left(\mathbf{x}_j^t,\boldsymbol{\boldsymbol{\xi}}^t\right)$ is obtained by optimizing the resource allocation given $\mathbf{x}_j^t$ in (P2). We will introduce the detailed algorithm to obtain $G\left(\mathbf{x}_j^t,\boldsymbol{\boldsymbol{\xi}}^t\right)$ in Section IV.B. Notice that the calculation of $G\left(\mathbf{x}_j^t,\boldsymbol{\boldsymbol{\xi}}^t\right)$ is performed by $M_t$ times to obtain the best action $\mathbf{x}^t$. Intuitively, a larger $M_t$ results in better solution performance, but a larger computation time. To balance the performance-complexity tradeoff, we propose here an adaptive procedure to set a time-varying $M_t$.

The key idea is that when the actor DNN gradually approaches the optimal policy over time, a small $M_t$ suffices to find the optimal action within a small distance to $\hat{\mathbf{x}}^t$. Denote $m_t \in [0,M_t-1]$ as the index of the best action $\mathbf{x}^t \in \Omega_t$. We define $m_t^* = \bmod(m_t, M_t/2)$, which represents the order of $\mathbf{x}^t$ among either the $M_t/2$ noise-free or the noise-added candidate actions. In practice, we set a maximum $M_1 =2N$ initially and update $M_t$ every $\delta_M \geq 1$ time frames. If $\bmod\left(t,\delta_M\right) = 0$ in time frame $t$, i.e., $t$ can be divided by $\delta_M$, we set
\begin{equation}
\label{108}
M_{t} = 2\cdot \min\left(\max \left(m_{t - 1}^*,\cdots, m_{t - \delta_M}^* \right)+1,N\right).
\end{equation}
The additional $1$ in the first term within the min operator allows $M_{t}$ to increase over time. Otherwise, $M_{t} = M_{t-1}$ if $\bmod\left(t,\delta_M\right) \neq 0$. Notice that too frequent update (small $\delta_M$) may degrade the training convergence while a too large $\delta_M$ cause unnecessary computational complexity.

\subsubsection{Policy Update Module} LyDROO uses $\left(\boldsymbol{\boldsymbol{\xi}}^t,\mathbf{x}^t\right)$ as a labeled input-output sample for updating the policy of the DNN. In particular, we maintain a replay memory that only stores the most recent $q$ data samples. In practice, with an initially empty memory, we start training the DNN after collecting more than $q/2$ data samples. Then, the DNN is trained periodically once every $\delta_T$ time slots to avoid model over-fitting. When $\bmod\left(t,\delta_T\right) = 0$, we randomly select a batch of data samples $\left\{\left(\boldsymbol{\boldsymbol{\xi}}^\tau,\mathbf{x}^\tau\right), \tau \in \mathcal{S}^t\right\}$, where $\mathcal{S}^t$ denotes the set of time indices of the selected samples. We then update the parameter of the DNN by minimizing its average cross-entropy loss function $LS\left(\boldsymbol{\theta}^t\right)$ over the data samples using the Adam algorithm \cite{marsland2015ml}
\begin{equation}
\begin{aligned}
  LS(\boldsymbol{\theta}^t) =& -1/|\mathcal{S}^t|\cdot {\mathsmaller \sum}_{\tau \in \mathcal{S}^t}\Big[ {(\mathbf{x}^{\tau})}^{\intercal} \log \Pi_{\boldsymbol{\theta}^t}\left(\boldsymbol{\boldsymbol{\xi}}^\tau\right) \\
  &\ \ \ \ \ \ \ \ \ \ \ \ + (1-\mathbf{x}^{\tau})^{\intercal}\log \big(1-\Pi_{\boldsymbol{\theta}^t}\left(\boldsymbol{\boldsymbol{\xi}}^\tau\right)\big)\Big],
   \end{aligned}
\end{equation}
where $|\mathcal{S}^t|$ denotes the size of the sample batch, $(\cdot)^\intercal$ denotes the transpose operator, and the log function denotes the element-wise logarithm operation of a vector. When the training completes, we update the parameter of the actor module in the next time frame to $\boldsymbol{\theta}^{t+1}$.

\begin{algorithm}
\scriptsize
 \SetAlgoLined
 \SetKwData{Left}{left}\SetKwData{This}{this}\SetKwData{Up}{up}
 \SetKwRepeat{doWhile}{do}{while}
 \SetKwFunction{Union}{Union}\SetKwFunction{FindCompress}{FindCompress}
 \SetKwInOut{Input}{input}\SetKwInOut{Output}{output}
 \Input{Parameters $V$, $\left\{\gamma_i, c_i\right\}_{i=1}^N$, $K$, training interval $\delta_T$, $M_t$ update interval $\delta_M$\;}
 \Output{Control actions $\left\{\mathbf{x}^t, \mathbf{y}^t\right\}_{t=1}^K$;}
 Initialize the DNN with random parameters $\boldsymbol{\theta}^{1}$ and empty replay memory, $M_1 \leftarrow 2N$\;
 Empty initial data queue $Q_i(1)=0$ and energy queue $Y_i(1)=0$, for $i=1,\cdots,N$\;
 \For{$t=1,2,\dots,K$}{
 Observe the input $\boldsymbol{\boldsymbol{\xi}}^t = \left\{h^t, Q_i(t), Y_i(t)\right\}_{i=1}^N$ and update $M_t$ using (\ref{108}) if $\bmod\left(t,\delta_M\right) = 0$\;
 Generate a relaxed offloading action $\hat{\mathbf{x}}^t = \Pi_{\boldsymbol{\theta}^t}\left(\boldsymbol{\boldsymbol{\xi}}^t\right)$ with the DNN\;
 Quantize $\hat{\mathbf{x}}_t$ into $M_t$ binary actions $\left\{\mathbf{x}^t_i| i = 1,\cdots,M_t\right\}$ using the NOP method\;
 Compute $G\left(\mathbf{x}^t_i, \boldsymbol{\boldsymbol{\xi}}^t\right)$ by optimizing resource allocation $\mathbf{y}^t_i$ in (P2) for each $\mathbf{x}^t_i$\; \label{step:compute_Q}
 Select the best solution $\mathbf{x}^{t}=\arg \underset{\{\mathbf{x}^t_i\}}{\max}\ G\left(\mathbf{x}^t_i, \boldsymbol{\boldsymbol{\xi}}^t\right)$ and execute the joint action $\left(\mathbf{x}^{t},\mathbf{y}^t\right)$\;
  Update the replay memory by adding $(\boldsymbol{\boldsymbol{\xi}}^t,\mathbf{x}^{t})$\;
 \If{$\bmod\left(t,\delta_T\right) = 0$}{
   Uniformly sample a batch of data set $\{( \boldsymbol{\boldsymbol{\xi}}^{\tau}, \mathbf{x}^{\tau})\mid \tau \in \mathcal{S}_t\}$ from the memory\;
   Train the DNN with $\{( \boldsymbol{\boldsymbol{\xi}}^{\tau}, \mathbf{x}^{\tau})\mid \tau \in \mathcal{S}_t\}$ and update $\boldsymbol{\theta}^{t}$ using the Adam algorithm\;
 }
 $t\leftarrow t+1$\;
 Update $\left\{Q_i(t), Y_i(t)\right\}_{i=1}^N$ based on $\left(\mathbf{x}^{t-1},\mathbf{y}^{t-1}\right)$ and data arrival observation $\left\{A^{t-1}_i\right\}_{i=1}^N$ using (\ref{111}) and (\ref{112}).
 }
 \caption{The online LyDROO algorithm for solving (P1).}
\end{algorithm}

\subsubsection{Queueing Module}
As a by-product of the critic module, we obtain the optimal resource allocation $\mathbf{y}^t$ associated with $\mathbf{x}^t$. Accordingly, the system executes the joint computation offloading and resource allocation action $\left\{\mathbf{x}^t,\mathbf{y}^t\right\}$, which processes data $\{D_i^t\}_{i=1}^N$ and consumes energy $\{e_i^t\}_{i=1}^N$ as given in (\ref{14}). Based on $\{D_i^t,e_i^t\}_{i=1}^N$ and the data arrivals $\{A_i^t\}_{i=1}^N$ observed in the $t$th time frame, the queueing module then updates the data and energy queues $\left\{Q_i(t+1), Y_i(t+1)\right\}_{i=1}^N$ using (\ref{111}) and (\ref{112}) at the beginning of the $(t+1)$th time frame. With the wireless channel gains observation $\{h_i^{t+1}\}_{i=1}^N$, the system feeds the input parameter $\boldsymbol{\boldsymbol{\xi}}^{t+1} = \left\{h_i^{t+1},Q_i(t+1), Y_i(t+1)\right\}_{i=1}^N$ to the DNN and starts a new iteration from the actor module in Step 1).

With the above actor-critic-update loop, the DNN consistently learns from the best and most recent state-action pairs, leading to a better policy $\pi_{\boldsymbol{\theta}^t}$ that gradually approximates the optimal mapping to solve (P3). We summarize the pseudo-code of LyDROO in Algorithm $1$, where the major computational complexity is in line $7$ that computes $G\left(\mathbf{x}^t_i, \boldsymbol{\boldsymbol{\xi}}^t\right)$ by solving the optimal resource allocation problems. This in fact indicates that the proposed LyDROO algorithm can be extended to solve (P1) when considering a general non-decreasing concave utility $U\left(r^t_i\right)$ in the objective, because the per-frame resource allocation problem to compute $G\left(\mathbf{x}^t_i, \boldsymbol{\boldsymbol{\xi}}^t\right)$ is a convex problem that can be efficiently solved, where the detailed analysis is omitted. In the next subsection, we propose a low-complexity algorithm to obtain $G\left(\mathbf{x}^t_i, \boldsymbol{\boldsymbol{\xi}}^t\right)$.

\subsection{Low-complexity Optimal Resource Allocation Algorithm}
Given the value of $\mathbf{x}^t$ in (P2), we denote the index set of users with $x^t_i=1$ as $\mathcal{M}_1^t$, and the complementary user set as $\mathcal{M}_0^t$. For simplicity of exposition, we drop the superscript $t$ and express the optimal resource allocation problem that computes $G\left(\mathbf{x}^t, \boldsymbol{\boldsymbol{\xi}}^t\right)$ as following
\begin{subequations}
   \label{8}
   \begin{align}
   (\textrm{P4}):\ & \underset{\boldsymbol{\tau},\mathbf{f},\mathbf{e}_O, \mathbf{r}_O}{\text{maximize}}\ \   \mathsmaller \sum_{j\in\mathcal{M}_0} \left\{a_j f_j/\phi -  Y_j(t) \kappa f_j^3 \right\} \notag \\
    & \ \ \ \ \ \ \ \ \ \ \ \ \ \ \ \  + \mathsmaller \sum_{i\in\mathcal{M}_1} \left\{a_i r_{i,O} - Y_i(t) e_{i,O} \right\} \notag\\
    & \text{subject to} \notag \\
    & f_j /\phi \leq  Q_j(t), 0 \leq f_j \leq f^{max}_j, \ \forall  j \in \mathcal{M}_0,  \notag\\
    & \mathsmaller \sum_{i\in \mathcal{M}_1} \tau_i  \leq 1, \notag \\
    & e_{i,O}\leq P^{max}_i \tau_i, \ r_{i,O} \leq  Q_i(t), \ \forall i \in  \mathcal{M}_1, \notag \\
    & r_{i,O} \leq \frac{ W\tau_i}{v_u}\log_2\left(1+\frac{e_{i,O} h_i}{\tau_i N_0}\right), \ \forall i \in  \mathcal{M}_1, \notag\\
    & \tau_i, r_{i,O},  e_{i,O} \geq 0, \ \forall i \in  \mathcal{M}_1, \notag
    \end{align}
\end{subequations}
where $a_i \triangleq Q_{i}(t) + V c_i$ is a parameter. Notice that (P4) can be separately optimized for WDs in $\mathcal{M}_1$ and $\mathcal{M}_0$. In particular, each $j\in \mathcal{M}_0$ solves an independent problem
\begin{equation}
\begin{aligned}
&\underset{f_j}{\text{maximize}}& &  a_j f_j/\phi -  Y_j(t) \kappa f_j^3\\
& \text{subject to} & & 0\leq  f_j \leq  \min\left\{\phi Q_j(t), f^{max}_j \right\},
\end{aligned}
\end{equation}
where the closed-form optimal solution is
\begin{equation}
\label{31}
f^{*}_j = \min \left\{ \sqrt{\frac{a_j}{3\phi \kappa Y_j(t)}}, \min\left\{\phi Q_j(t), f^{max}_j \right\} \right\}, \ \forall j\in \mathcal{M}_0.
\end{equation}
Intuitively, the $j$th WD computes faster when $Q_j(t)$ is large or $Y_j(t)$ is small, and vice versa.

On the other hand, denote $\boldsymbol{\hat{\tau}} = \left\{\tau_i, \forall i\in \mathcal{M}_1\right\}$, $\mathbf{\hat{e}}_O = \left\{e_{i,O}, \forall i\in \mathcal{M}_1\right\}$ and $\mathbf{\hat{r}}_O = \left\{r_{i,O}, \forall i\in \mathcal{M}_1\right\}$, we need to solve the following problem for the WDs in $\mathcal{M}_1$,
\begin{subequations}
    \label{37}
   \begin{align}
    & \underset{\boldsymbol{\hat{\tau}}, \mathbf{\hat{e}}_O, \mathbf{\hat{r}}_O}{\text{maximize}} & &  \mathsmaller \sum_{i\in\mathcal{M}_1} \left\{a_i r_{i,O} - Y_i(t) e_{i,O} \right\} \notag \\
    & \text{subject to} &  & \mathsmaller \sum_{i\in \mathcal{M}_1} \tau_i  \leq 1, \notag \\
    & & &   e_{i,O}\leq P^{max}_i \tau_i, r_{i,O} \leq  Q_i(t), \ \forall i \in \mathcal{M}_1, \notag\\
    & & &   r_{i,O} \leq \frac{ W\tau_i}{v_u}\log_2\left(1+\frac{e_{i,O} h_i}{\tau_i N_0}\right), \ \forall i \in  \mathcal{M}_1. \notag\\
    & & &  \tau_i, r_{i,O},  e_{i,O} \geq 0, \ \forall i \in  \mathcal{M}_1. \notag
    \end{align}
\end{subequations}
We express a partial Lagrangian of the problem as
\begin{equation}
    \label{36}
   \begin{aligned}
    &L\left(\left\{\boldsymbol{\hat{\tau}}, \mathbf{\hat{e}}_O, \mathbf{\hat{r}}_O\right\}, \mu \right) \\
    =&  \mathsmaller \sum_{i\in\mathcal{M}_1} \left\{a_i r_{i,O} - Y_i(t) e_{i,O} \right\} + \mu \left(1- \mathsmaller \sum_{i\in \mathcal{M}_1} \tau_i\right),
    \end{aligned}
\end{equation}
where $\mu$ denotes the dual variable. Furthermore, the dual function is
\begin{subequations}
\label{17}
\begin{align}
d(\mu) =\  & \underset{\boldsymbol{\hat{\tau}}, \mathbf{\hat{e}}_O, \mathbf{\hat{r}}_O}{\text{maximize}} \ \  L\left(\left\{\boldsymbol{\hat{\tau}}, \mathbf{\hat{e}}_O, \mathbf{\hat{r}}_O\right\}, \mu \right) \notag\\
& \text{subject to}  \notag\\
& e_{i,O}\leq P^{max}_i\tau_i, \ r_{i,O}\leq Q_i(t),\ \forall  i \in \mathcal{M}_1,  \notag \\
& r_{i,O} \leq \frac{ W\tau_i}{v_u}\log_2\left(1+\frac{e_{i,O} h_i}{\tau_i N_0}\right), \ \forall i \in  \mathcal{M}_1, \notag\\
& \tau_i, r_{i,O}, e_{i,O} \geq 0,  \ \forall  i \in \mathcal{M}_1 \notag
\end{align}
\end{subequations}
and the dual problem is $\underset{\mu\geq 0}{\text{minimize}}\ d(\mu)$. Notice that the dual function can be decomposed into parallel sub-problems. For a WD $i\in \mathcal{M}_1$, it solves
\begin{subequations}
\label{90}
\begin{align}
& \underset{\tau_i,e_{i,O},r_{i,O}}{\text{maximize}} & &  \left\{a_i r_{i,O} - Y_i(t) e_{i,O} \right\} - \mu \tau_i \\
& \text{subject to} &  &  \tau_i \geq 0, 0 \leq e_{i,O}\leq P^{max}_i\tau_i, \label{92} \\
& & & 0 \leq r_{i,O}\leq Q_i(t),\\
& & & \ r_{i,O} \leq \frac{ W\tau_i}{v_u}\log_2\left(1+\frac{e_{i,O} h_i}{\tau_i N_0}\right). \label{89}
\end{align}
\end{subequations}
In the following, we propose a simple algorithm that solves (\ref{90}) efficiently.

Notice that equality (\ref{89}) holds at the optimum because otherwise we can reduce the value of $e_{i,O}$ at the optimum to achieve a higher objective. By setting $r_{i,O} = \frac{ W\tau_i}{v_u}\log_2\left(1+\frac{e_{i,O} h_i}{\tau_i N_0}\right)$ in (\ref{89}), we can equivalently write the constraint $0 \leq  e_{i,O}\leq P^{max}_i\tau_i$ in (\ref{92}) as
\begin{equation}
\label{35}
0 \leq \frac{r_{i,O}}{\tau_i} \leq \frac{ W }{v_u}\log_2\left(1+\frac{P_i^{max} h_i}{ N_0}\right) \triangleq  R_i^{max},
\end{equation}
where $R_i^{max}$ denotes the maximum transmission rate of the $i$th WD. From (\ref{14}), we express $e_{i,O}$ as a function of $r_{i,O}$ and $\tau_i$ as
\begin{equation}
\label{32}
e_{i,O} = \frac{N_0 \tau_i}{h_i} \left(2^{\frac{r_{i,O} v_u}{W \tau_i}}-1\right) \triangleq \frac{\tau_i}{h_i}g\left(\frac{r_{i,O}}{\tau_i}\right),
\end{equation}
where $g(x) \triangleq N_0 \left(2^{\frac{x v_u}{W}}-1\right)$ is a convex function. By plugging (\ref{35}) and (\ref{32}) into (\ref{90}), we can equivalently transform (\ref{90}) as the following problem
\begin{equation}
\label{18}
\begin{aligned}
&\underset{r_{i,O},\tau_i}{\text{maximize}} & &  a_i r_{i,O} -\mu \tau_i - Y_i(t) \frac{\tau_i}{h_i}g\left(\frac{r_{i,O}}{\tau_i}\right)  \\
& \text{subject to} &  &  \tau_i \geq  r_{i,O}/R_i^{max},\ 0 \leq r_{i,O}\leq Q_i(t).
\end{aligned}
\end{equation}
Notice that (\ref{18}), and thus (\ref{90}), is equivalent to the following problem
\begin{equation}
\label{33}
\underset{r_{i,O}}{\text{maximize}} \left\{V_i(r_{i,O})| 0\leq r_{i,O} \leq Q_i(t)\right\},
\end{equation}
where
\begin{equation}
\label{20}
\begin{aligned}
V_i(r_{i,O}) \triangleq\  &\underset{\tau_i}{\text{maximize}} & &  a_i r_{i,O} -\mu \tau_i - Y_i(t) \frac{\tau_i}{h_i}g\left(\frac{r_{i,O}}{\tau_i}\right)  \\
& \text{subject to} &  &  \tau_i \geq  r_{i,O}/R_i^{max}.
\end{aligned}
\end{equation}
(\ref{20}) is a convex problem, where we derive the optimal solution in the following Proposition $1$.

\textbf{Proposition 1:} The optimal solution of (\ref{20}) is
\begin{equation}
\label{28}
\tau_i^* = \begin{cases}
\frac{r_{i,O}}{R_i^{max}},&   \text{if\ } h_i \leq \psi_i(\mu),\\
\frac{\ln 2 v_u \cdot r_{i,O}}{W\cdot \left[  \mathcal{W}\left(e^{-1}\left[\frac{\mu h_i}{Y_i(t) N_0} -1\right]\right)+ 1\right]},   &   \text{otherwise},\\
\end{cases}
\end{equation}
where $\psi_i(\mu) \triangleq \frac{ N_0}{P_i^{max}}\left(\frac{A_i}{-\mathcal{W}\left( - A_i\exp\left(-A_i\right) \right)}-1\right)$ and $A_i \triangleq 1+ \frac{\mu}{Y_i(t)P_{max}}$ are fixed parameters given $\mu$. $\mathcal{W}(x)$ denotes the Lambert-W function, which is the inverse function of $J(z) = z \exp(z) =x$, i.e., $z = \mathcal{W}(x)$.

\emph{Proof}: Please see the detailed proof in the Appendix A. $\hfill \blacksquare$

\emph{Remark 2:} A close observation of (\ref{28}) shows that we can compactly express the optimal solution $\tau_i^*$ of problem (\ref{20}) as $r_{i,O} = l_i(\mu) \tau_i^*$, where $l_i(\mu)$ is a fixed parameter given $\mu$, representing the optimal communication data rate of the $i$th WD. In other words, the optimal transmission time $\tau_i^*$ of (\ref{20}) increases linearly with $r_{i,O}$ under a fixed transmission rate $l_i(\mu)$. In the following, we show how to obtain the dual optimal solution $\mu^*$ and retrieve the primal optimal solutions to (\ref{37}) accordingly.

From Proposition $1$, we plug $r_{i,O} = l_i(\mu) \tau_i$ into (\ref{20}) and rewrite problem (\ref{33}) as
\begin{equation}
\begin{aligned}
&\underset{r_{i,O}}{\text{maximize}} & &  \left\{a_i - \frac{\mu}{l_i(\mu)}- Y_i(t) \frac{g\left[l_i(\mu)\right]}{l_i(\mu)h_i} \right\}r_{i,O}  \\
& \text{subject to} &  &  0 \leq r_{i,O} \leq Q_i(t),
\end{aligned}
\end{equation}
where the optimal solution is
\begin{equation}
\label{34}
r_{i,O}^* = \begin{cases}
0,&   \text{if\ } a_i - \frac{\mu}{l_i(\mu)}- Y_i(t) \frac{g\left[l_i(\mu)\right]}{l_i(\mu)h_i} < 0,\\
Q_i(t),   &   \text{otherwise}.\\
\end{cases}
\end{equation}
Accordingly, we have $\tau_i^* = r_{i,O}^*/l_i(\mu)$. After obtaining $\tau_i^*$, $\forall i\in \mathcal{M}_1$, we calculate the subgradient of $\mu$ in (\ref{36}) as $1- \sum_{i\in \mathcal{M}_1} \tau_i^*$. Then, we obtain the optimal dual variable $\mu^*$ through the ellipsoid method (bi-section search in this case) over the range $[0, \Delta]$, where $\Delta$ is a sufficiently large value, until a prescribed precision requirement is met.

Given the optimal $\mu^*$, we denote the optimal ratio obtained from (\ref{28}) as $l_i\left(\mu^*\right) \triangleq r_{i,O}^*/\tau_i^*$, $\forall i\in \mathcal{M}_1$. Notice that the optimal solution $\left\{\tau_i^*,r_{i,O}^*, \forall i \in \mathcal{M}_1\right\}$ of the dual problem may not be primal feasible. Therefore, to find a primal optimal solution to (\ref{37}), we substitute $\tau_i  = r_{i,O}/l_i\left(\mu^*\right)$ into (\ref{37}) and simplify the problem as
\begin{equation}
    \label{38}
   \begin{aligned}
    & \underset{\mathbf{\hat{r}}_O}{\text{maximize}} & &  \sum_{i\in\mathcal{M}_1} \left\{a_i - \frac{Y_i(t) g\left[l_i(\mu^*)\right]}{h_i l_i(\mu^*)} \right\} r_{i,O} \\
    & \text{subject to} &  & \sum_{i\in \mathcal{M}_1} \frac{r_{i,O}}{l_i(\mu^*)} \leq 1, \  r_{i,O} \leq  Q_i(t) , \forall i \in \mathcal{M}_1.
    \end{aligned}
\end{equation}
The above problem is a simple linear programming (LP) that can be easily solved. With a bit abuse of notation, we denote the optimal solution of (\ref{38}) as $\mathbf{\hat{r}}_O^* = \left\{r_{i,O}^*, \forall i\in \mathcal{M}_1\right\}$ and retrieve the optimal solution to (\ref{37}) as
\begin{equation}
\label{39}
\tau_i^* = r^*_{i,O} / l_i\left(\mu^*\right), \ e_{i,O}^* = \frac{\tau_i^* g\left[l_i(\mu^*)\right]}{h_i l_i(\mu^*)}, \ \forall i\in \mathcal{M}_1.
\end{equation}
Denote $\boldsymbol{\hat{\tau}}^* = \left\{\tau_i^*, \forall i\in \mathcal{M}_1\right\}$ and $\mathbf{\hat{e}}^*_O = \left\{e_{i,O}^*, \forall i\in \mathcal{M}_1\right\}$. As $\left\{\boldsymbol{\hat{\tau}}^*,\mathbf{\hat{e}}^*_O,\mathbf{\hat{r}}_O^*,\mu^*\right\}$ satisfies the KKT conditions, $\left\{\boldsymbol{\hat{\tau}}^*,\mathbf{\hat{e}}^*_O,\mathbf{\hat{r}}_O^*\right\}$ is an optimal solution to (\ref{37}). By combining the optimal solutions in (\ref{31}) and (\ref{39}), we obtain an optimal solution of (P4). We summarize the pseudo-code of the algorithm to solve (P4) in Algorithm $2$.

\section{Performance Analysis}
\subsection{Computational Complexity}
We first analyze the complexity of the proposed LyDROO scheme. The execution of LyDROO algorithm consists of two parts: offloading action generation (line 4-9 of Algorithm $1$) and policy update (line 10-13 of Algorithm $1$). In between, offloading action generation needs to be performed in every time frame, while policy update is performed infrequently (e.g., once every tens of time frames) and in parallel with task offloading and local computation. Therefore, we focus on analyzing the complexity of generating an offloading action in each time frame. A close observation shows that the major complexity is on optimizing the resource allocation in line $7$ of Algorithm $1$, which executes Algorithm $2$ to solve (P4) $M_t$ times in each time frame.

\begin{algorithm}
\scriptsize
 \SetAlgoLined
 \SetKwData{Left}{left}\SetKwData{This}{this}\SetKwData{Up}{up}
 \SetKwRepeat{doWhile}{do}{while}
 \SetKwFunction{Union}{Union}\SetKwFunction{FindCompress}{FindCompress}
 \SetKwInOut{Input}{input}\SetKwInOut{Output}{output}
  \Input{$\mathbf{x}^t$, $\boldsymbol{\boldsymbol{\xi}}^t = \left\{Y_i(t),Q_i(t),A_{i}^t\right\}_{i=1}^N$}
 \textbf{initialization:}  $\sigma_0 \leftarrow 0.1$, $LB \leftarrow 0$, $UB \leftarrow$ sufficiently large value, convert $\mathbf{x}^t$ into $\{\mathcal{M}_0,\mathcal{M}_1\}$ in (P4)\;
     \For{\emph{each WD $j\in \mathcal{M}_0$}}{
      Calculate $f^*_j$ using (\ref{31})\;
      }

      \Repeat{$|UB-LB|\leq \sigma_0$}{
      $\mu \leftarrow \frac{UB+LB}{2}$\;

      \For{\emph{each WD $i\in \mathcal{M}_1$}}{
      Calculate $l_i(\mu)$ using (\ref{28}) and $r_{i,O}^*$ using (\ref{34})\;
      $\tau_i^* \leftarrow r_{i,O}^*/l_i(\mu)$\;
 }
 \eIf{$1-\sum_{i\in \mathcal{M}_1} \tau_i^*<0$}{
      $LB \leftarrow \mu$\;
      }{
      $UB \leftarrow \mu$\;
      }
 }
 $\mu^*\leftarrow \mu$ and obtain $\mathbf{\hat{r}}_O^*$ by solving the LP in (\ref{38}), then obtain $\boldsymbol{\hat{\tau}}^*$ and $\mathbf{\hat{e}}_O^*$ using (\ref{39})\;
 \textbf{Return} an optimal solution of (P4) by combining (\ref{31}) and (\ref{39}).
 \caption{Primal dual algorithm for optimal resource allocation of (P4)}
\end{algorithm}

We show that the time complexity of Algorithm $2$ is $O\left(N\log_2\left(\frac{\Delta}{\sigma_0}\right)+N^3 \bar{L}\right)$: the first term corresponds to the bi-section search of $\mu$ with $\sigma_0$ being the small positive precision parameter; the second term corresponds to solving the LP in (\ref{38}) using interior point method \cite{1999:Anstreicher} with $\bar{L}$ being the length of input in binary representation to problem (\ref{38}). Compared to directly solving a general convex optimization (P4) with $4N$ variables using the interior point method, the proposed Algorithm $2$ solves an LP in (\ref{38}) with only $N$ variables, which incurs much lower computational complexity especially when $N$ is large. Since LyDROO executes Algorithm $2$ for $M_t$ times in each time frame, the overall complexity of generating an offloading action is $O\left(\left[N\log_2\left(\frac{\Delta}{\sigma_0}\right)+N^3 \bar{L}\right] M_t\right)$. Thanks to the adaptive procedure in (\ref{108}) that gradually reduces the value of $M_t$ during the learning process, we observe in simulations that a small $M_t$ (e.g., less than $5$ when $N=30$) suffices to generate optimal offloading action when the learning process converges. In the Section VI, we show in simulations that the proposed LyDROO enjoys very low computation time and is suitable for online implementation in time-varying edge environment.

\subsection{Convergence Performance}
We then analyze the asymptotic convergence performance of the LyDROO algorithm in solving (P1). To begin with, we first introduce some preliminaries of Lyapunov optimization. We denote the \emph{random event} of the considered problem as an i.i.d. process $\omega(t)$, which consists of the fading channels and data arrivals, i.e., $\omega(t) = \left\{h^t_i,A^t_i\right\}_{i=1}^N$. We introduce a class of stationary and randomized policies called \emph{$\omega$-only policy}, which observes $\omega(t)$ for each time frame $t$ and makes control decisions independent of the queue backlog $\mathbf{Z}(t)$. To ensure that the data queue stability constraint can be satisfied, we assume (P1) is feasible and the following Slater condition holds.

\textbf{Assumption 1 (Slater Condition)}: There are values $\epsilon>0$ and $\Phi\left(\epsilon\right) \leq R^{opt}$ and a $\omega$-only policy $\Pi$ that makes control decisions $\alpha^{\Pi,t}$ in the $t$th time frame, which satisfy
\begin{equation}
\allowdisplaybreaks
 \label{81}
 \begin{aligned}
&\mathbb{E}\left[R^t\left(\alpha^{\Pi,t}\right)\right] = \Phi\left(\epsilon\right),\\
&\mathbb{E}\left[e_i^t\left(\alpha^{\Pi,t}\right) \right] \leq   \gamma_i -\epsilon, \ \forall i. \\
&\mathbb{E}\left[A_i^t\right] \leq \mathbb{E}\left[D_i^t\left(\alpha^{\Pi,t}\right)\right] - \epsilon, \ \forall i.
\end{aligned}
\end{equation}
Here, $R^t \triangleq \sum_{i=1}^N c_i r_i^t$ denotes the weighted sum computation rate archived in the $t$th time frame. $R^{opt}$ is the optimal objective of (P1) obtained over all feasible control policies (including but not limited to $\omega$-only policy).

We show the performance of LyDROO algorithm in the following Theorem $1$.

\textbf{Theorem 1}: Suppose that (P1) is feasible and satisfies the Slater condition for some $\epsilon$, $\Phi\left(\epsilon\right)$ and $\omega$-only policy $\Pi$. Suppose that given any $\mathbf{Z}(t)$ in time frame $t$, the LyDROO algorithm produces a value of (\ref{12}) that is no larger than a constant $C\geq 0$ above the minimum, i.e., the per-frame subproblem (P2) is solved within an optimality gap $C$. Then, the following conditions hold when applying the LyDROO algorithm in each time frame $t$
\begin{enumerate}
  \item[a)] The time average computation rate satisfies
  \begin{equation}
\lim_{K\rightarrow \infty} 1/K \cdot \mathsmaller \sum_{t=1}^{K}  \mathbb{E}\left[R^t\right] \geq  R^{opt} - (\hat{B}+C)/V.
\end{equation}
  \item[b)] The average system queue length satisfies
    \begin{equation}
    \label{76}
    \begin{aligned}
&\lim_{K\rightarrow \infty} 1/K \cdot \mathsmaller\sum_{t=1}^{K}  \mathsmaller\sum_{i=1}^N \mathbb{E}\left[Q_i^t\right]  \\
&\leq  1/\epsilon \cdot \left(\hat{B}+C + V \left[ R^{opt}-\Phi\left(\epsilon\right)\right]\right).
\end{aligned}
\end{equation}
  \item[c)] All the data queues $Q_i(t)$ are strongly stable and the time average power constraint (\ref{63}) is satisfied with probability $1$.
\end{enumerate}

\emph{Proof}: Please see the detailed proof in the Appendix B. $\hfill \blacksquare$

Theorem $1$ indicates that if the LyDROO algorithm achieves a limited optimality gap $C$ when solving the per-frame subproblem (P2), then we satisfy all the long-term constraints and achieve an $\left[O(1/V),O(V)\right]$ computation rate-delay tradeoff. That is, by increasing $V$, we can improve the objective of (P1) proportional to $1/V$, but at the cost of longer data queue length (processing delay) proportional to $V$, and vice versa. Besides, a smaller $C$ leads to both higher rate and lower delay performance. In simulation section, we demonstrate the impact of $V$ on the long-term performance and show that LyDROO achieves very small $C$ for the per-frame subproblem. Notice that the above analysis does not assume the specific utility function in the objective of (P1), and thus the results hold for any non-decreasing concave utility function $U(r_i^t)$.

\section{Simulation Results}
In this section, we use simulations to evaluate the performance of the proposed LyDROO algorithm.\footnote{The source code is available at https://github.com/revenol/LyDROO.} All the computations are evaluated on a TensorFlow 2.0 platform with an Intel Core i5-4570 3.2GHz CPU and 12 GB of memory. We assume that the average channel gain $\bar{h}_i$ follows a path-loss model $\bar{h}_i = A_d\left(\frac{3\times 10^8}{4\pi f_c d_i}\right)^{d_e}$, $i=1,\cdots,N$, where $A_d =3$ denotes the antenna gain, $f_c =915$ MHz denotes the carrier frequency, $d_e =3$ denotes the path loss exponent, and $d_i$ in meters denotes the distance between the $i$th WD and the ES. $h_i$ follows an i.i.d. Rician distribution with line-of-sight link gain equal to $0.3 \bar{h}_i$. The noise power $N_0 = W \upsilon_0$ with noise power spectral density $\upsilon_0 = -174$ dBm/Hz. Unless otherwise stated, we consider $N=10$ WDs equally spaced with $d_i = 120 + 15(i-1)$, for $i=1,\cdots,N$. The weight $c_i=1.5$ if $i$ is an odd number and $c_i=1$ otherwise. The task data arrivals of all the WDs follow exponential distribution with equal average rate $\mathbb{E}\left[A_{i}^t\right] = \lambda_i$, $i=1,\cdots,N$. The values of the other parameters are listed in Table \ref{tab:parameter}, which are equal for all the WDs.

\begin{table}
\caption{Simulation Parameters}
\footnotesize
\begin{center}
\begin{tabular}{|l|| l|| l| }
\hline
  $W = 2$ MHz                  & $f_i^{max} = 0.3$ GHz     &   $P_i^{max} = 0.1$ watt     \\ \hline
  $v_u = 1.1$                  &   $\kappa = 10^{-26}$     &   $\phi = 100$              \\ \hline
  $q = 1024$                   &   $\delta_T = 10$         &   $\delta_M=32$             \\ \hline
  $\gamma_i = 0.08$ watt       &   $\nu = 1000$            &   $|\mathcal{S}^t|= 32$    \\ \hline
  $V=20$                       &   $\lambda_i = 3$ Mbps    &   $T=1$ second                \\ \hline
\end{tabular}
\end{center}
\label{tab:parameter}
\end{table}

The proposed LyDROO adopts a fully-connected multilayer perceptron in the actor module, consisting of one input layer, two hidden layers, and one output layer, where the first and second hidden layers have $120$ and $80$ hidden neurons, respectively. For performance comparison, we consider two benchmark methods:
\begin{itemize}
  \item Lyapunov-guided Coordinate Decent (LyCD): It minimizes the upper bound of drift-plus-penalty in (\ref{12}), or equivalently solves (P2), using the coordinate decent (CD) method \cite{2018:Bi} that iteratively applies one-dimensional search to update the binary offloading decision vector $\mathbf{x}^t$. Although the optimal solution of (P2) is hard to obtain, we have verified through extensive simulations that the CD method achieves close-to-optimal performance. Therefore, we consider LyCD as a target performance benchmark of the LyDROO algorithm. The major drawback of LyCD, however, lies in the significant computation delay when $N$ is large. We show in the following simulations that the proposed LyDROO achieves the similar computation performance as LyCD but takes much shorter computation time.
  \item Myopic optimization \cite{2019:Huang}: The Myopic method neglects the data queue backlogs and maximizes the weighted sum computation rate in each time frame $t$ by solving
    \begin{subequations}
   \begin{align}
    & \underset{\mathbf{x}^t, \boldsymbol{\tau}^t,\mathbf{f}^t,\mathbf{e}_O^t, \mathbf{r}_{O}^t}{\text{maximize}} & &   \mathsmaller\sum_{i=1}^N c_i r^{t}_{i} \\
    & \text{subject to} &  & (\ref{25})-(\ref{85}), \\
    &                  &   & e_i^t \leq t\gamma_i - \mathsmaller\sum_{l=1}^{t-1} e_i^l, \ \forall i. \label{97}
   \end{align}
    \end{subequations}
Here, constraint (\ref{97}) guarantees that the $i$th average power constraint of (P1) is satisfied up to the $t$th time frame, where $\left\{e_i^l | l<t\right\}$ is the past energy consumptions known at the $t$th time frame.
\end{itemize}
Besides the two benchmarks above, we have also considered using Deep Deterministic Policy Gradient (DDPG) \cite{2016:Lillicrap}, a state-of-the-art policy-based DRL scheme, to directly learn the optimal mapping from the input $\mathbf{\xi}^t$ to the output mixed integer-continuous offloading action $\{x_i^t,\tau_i^t, f_i^t, e_{i,O}^t\}_{i=1}^N$ in (P2). However, we find through extensive simulations that DDPG is unable to stabilize the task data queues for all the WDs, even when the number of WDs and task arrival rates are small, e.g., $N=3$ and $\lambda_i=3$ Mbps. Therefore, we do not include DDPG as a performance benchmark in the following simulations.

\begin{figure}
\centering
\includegraphics[width=0.48\textwidth]{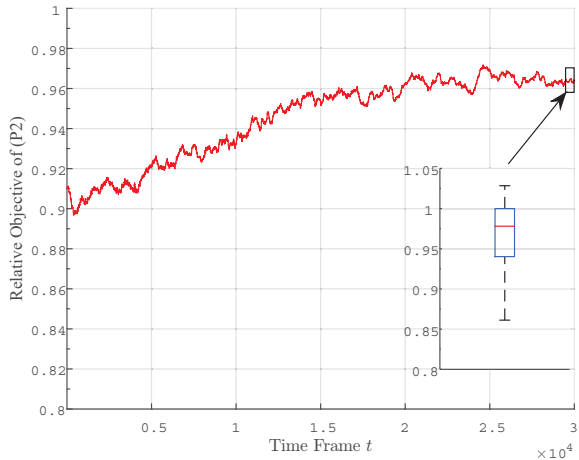}
\caption{Performance of the LyDROO algorithm in solving per-frame subproblem (P2). In the boxplot, the central mark (in red) indicates the median, and the bottom and top edges of the box indicate the $25$th and $75$th percentiles, respectively.}
\label{104}
\end{figure}

In Fig.~\ref{104}, we first evaluate the performance of the LyDROO algorithm in solving per-frame subproblem (P2). For fair comparison, we first apply the LyCD method for $30,000$ time frames, where we record the input to the actor module $\left\{\boldsymbol{\boldsymbol{\xi}}(t)\right\}$ throughout the time. Then, we use the same $\left\{\boldsymbol{\boldsymbol{\xi}}(t)\right\}$ as the input to the LyDROO framework in Fig.~\ref{fig:network} only for computing the output action $\left\{\mathbf{x}^t,\mathbf{y}^t\right\}$ in each time frame without updating the queue states. We plot the ratio between the objective values of (P2) achieved by the LyDROO and LyCD as the time proceeds, where each point is a moving-window average of $500$ time frames. We notice that the ratio gradually increases with time and eventually reaches about $0.96$. We also show the boxplot of the last $500$ time frames, which shows that the medium ratio is around $0.98$ and the ratio is above $0.94$ for more than $75\%$ of the cases. As LyCD achieves close-to-optimal performance of the per-frame subproblem (P2), this shows that LyDROO solves (P2) with small optimality gap $C$, thus leading to both higher computation rate and lower execution delay according to Theorem $1$.

\begin{figure}
  \centering
  \subfigure{\includegraphics[width=0.45\textwidth]{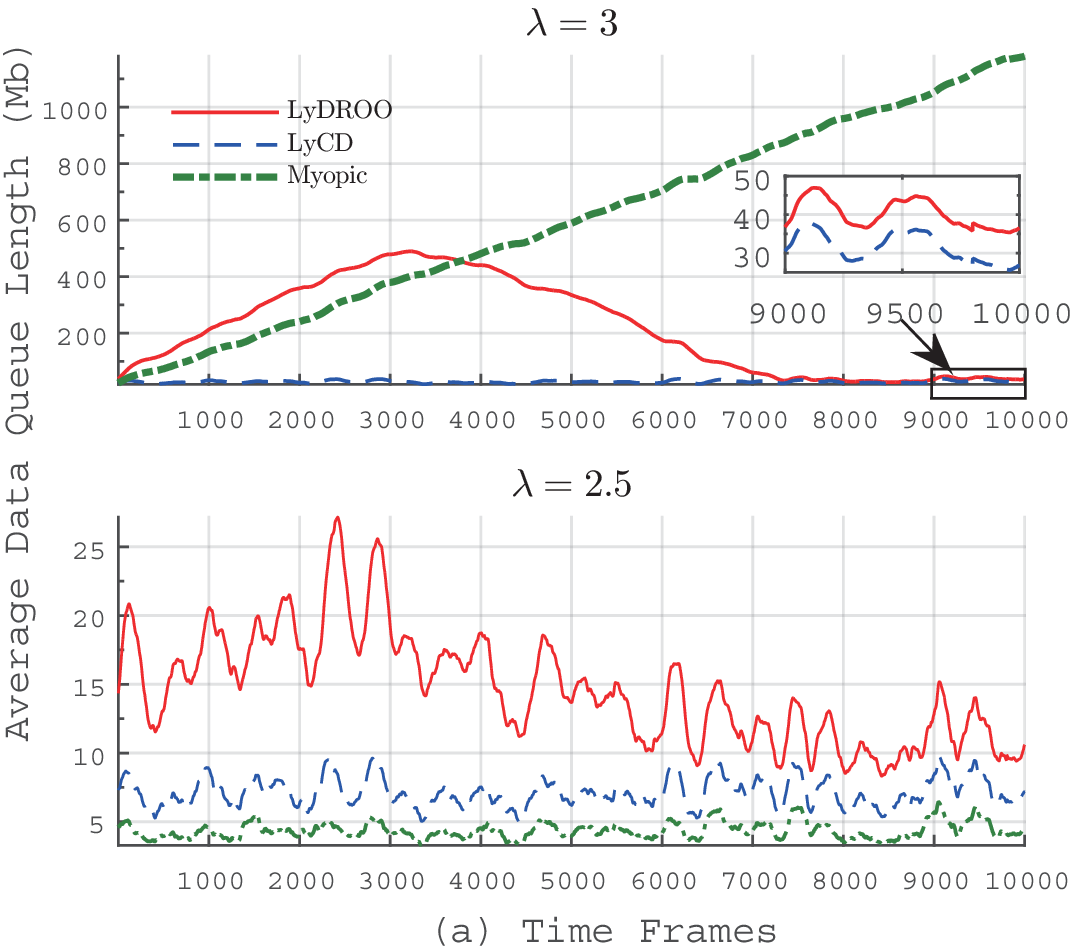}}\quad
  \subfigure{\includegraphics[width=0.47\textwidth]{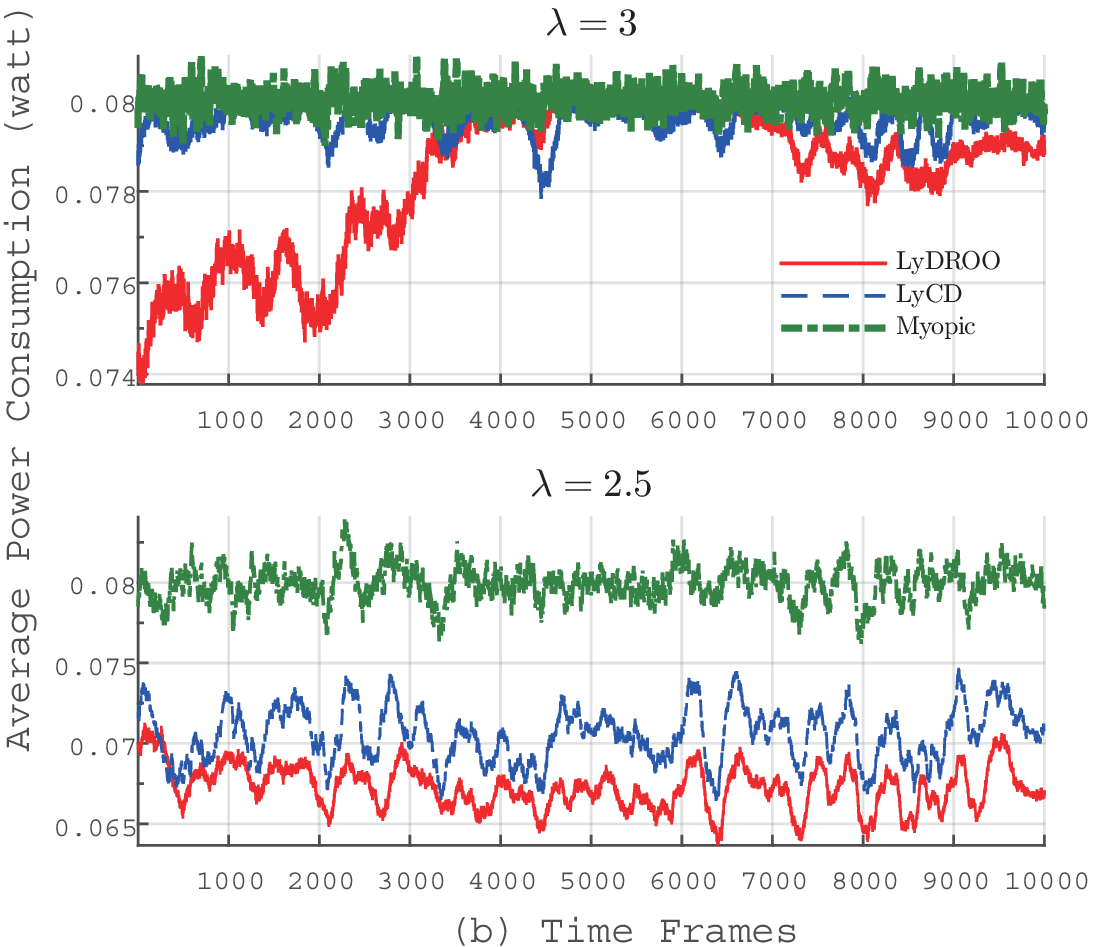}}\quad
  \subfigure{\includegraphics[width=0.45\textwidth]{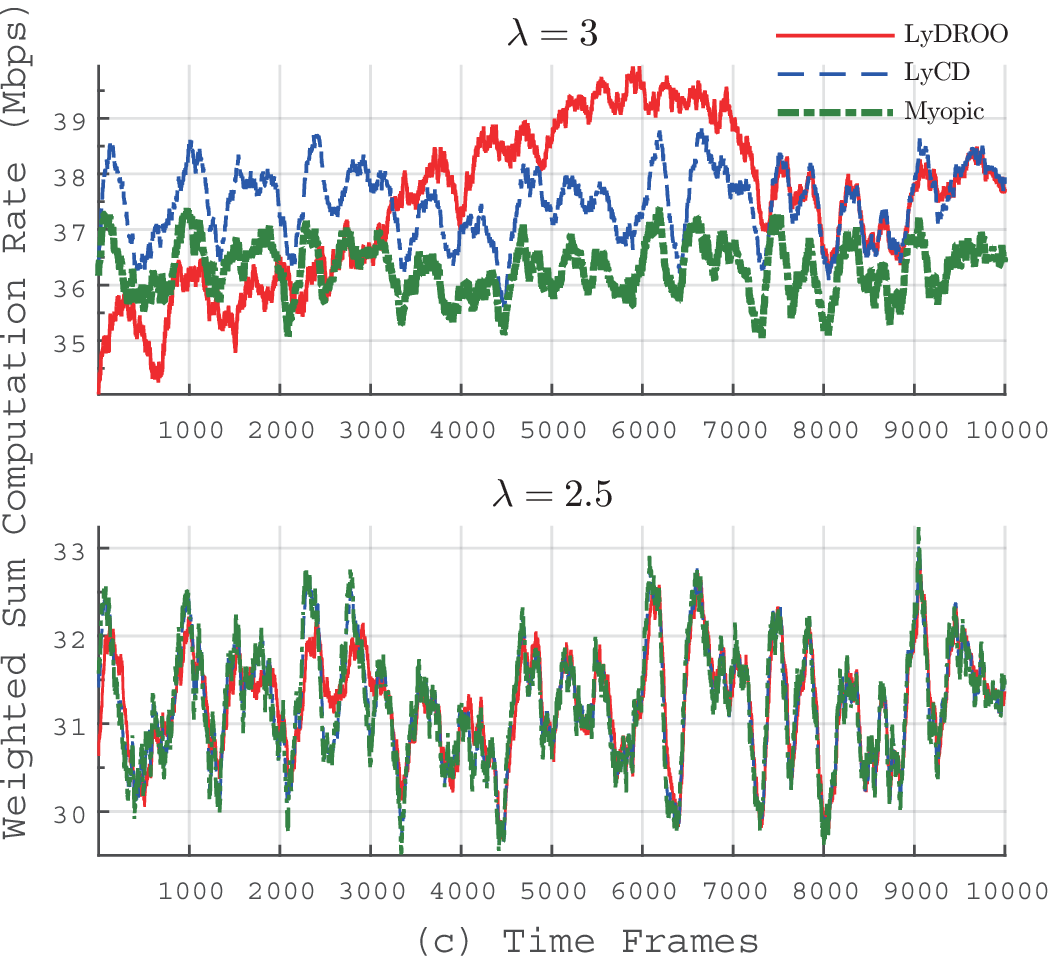}}
  \caption{Convergence performance comparisons of different schemes under $\lambda_i = 2.5$ and $3$. From top to bottom figures: a) data queue length; b) power consumption; c) weighted sum computation rate.}
  \label{103}
\end{figure}

We then evaluate the convergence of proposed LyDROO algorithm and the two benchmark methods. In Fig.~\ref{103}, we consider two data arrival rates with $\lambda_i = 2.5$ and $3$ Mbps for all $i$, and plot the weighted sum computation rate, average data queue length, and average power consumption performance over time. We consider i.i.d. realizations of random events $\omega(t)$ in $10,000$ time frames, where each point in the figure is a moving-window average of $200$ time frames. In Fig.~\ref{103}(a), we observe that for a low data arrival rate $\lambda_i=2.5$, all the schemes maintain the data queues stable and achieve similar computation rate performance. Besides, they all satisfy the average power constraint $0.08$ W in Fig.~\ref{103}(b). In particular, the LyDROO and LyCD methods achieve higher data queue lengths than the Myopic scheme, as they consume strictly lower power than the average power requirement, meanwhile achieving the identical rate performance in Fig.~\ref{103}(c). When we increase $\lambda_i$ from $2.5$ to $3$, all the three schemes still satisfy the average power constraints. However, the average data queue length of the Myopic method increases almost linearly with time, indicating that it is unable to stabilize the data queues. This is because the data arrival rate has surpassed the computation capacity (i.e., achievable sum computation rate) of the Myopic algorithm. On the other hand, both the LyCD and LyDROO methods can stabilize the data queues, indicating that the proposed Lyapunov-based method can achieve a higher computation capacity than the Myopic method. In between, the LyCD method maintains lower data queue length over all time frames. The LyDROO method takes time to learn the optimal offloading policy in the early stage, where the data queue length increases quickly when $t\leq 3,000$. However, as the embedded DNN gradually approaches the optimal policy, the data queue length quickly drops and eventually converges to the similar queue length and rate performance as the LyCD method after around $t=7,500$. For both $\lambda_i$'s, the data queue lengths of the LyDROO algorithm start to drop at around $t=3,000$, indicating its fast convergence even under highly dynamic queueing systems. We also notice that the proposed LyDROO achieves excellent computation performance even before the learning algorithm fully converges. In Fig.~\ref{103}(c) with $\lambda_i=3$ Mbps, the performance gap of computation rate is less than $5\%$ when $t\leq 3000$ compared to the target benchmark LyCD, and LyDROO even achieves higher rate than LyCD between $t=3000$ to $7000$ when the learning process gradually converges.

\begin{figure}
  \centering
  \subfigure{\includegraphics[width=0.45\textwidth]{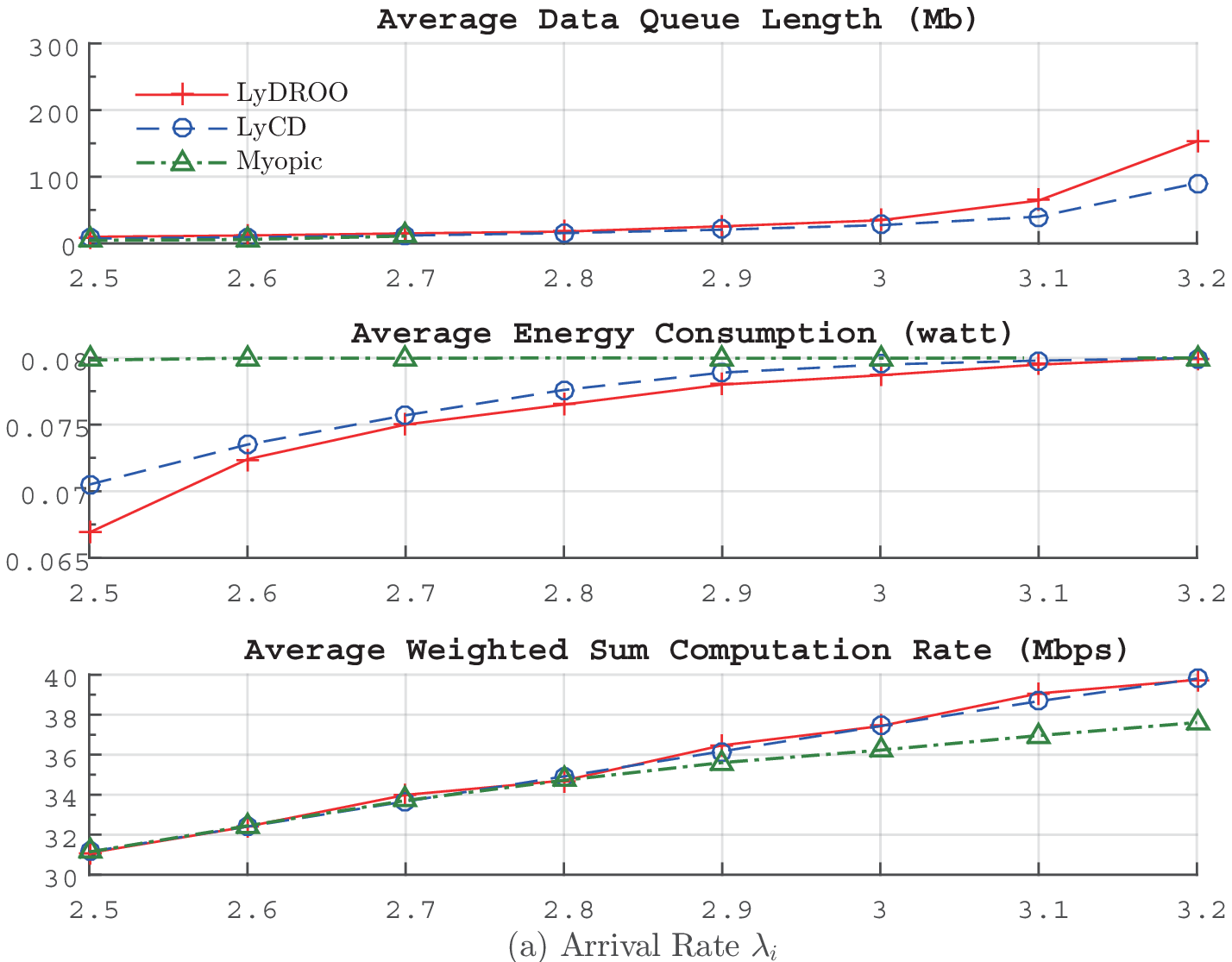}}\quad
  \subfigure{\includegraphics[width=0.45\textwidth]{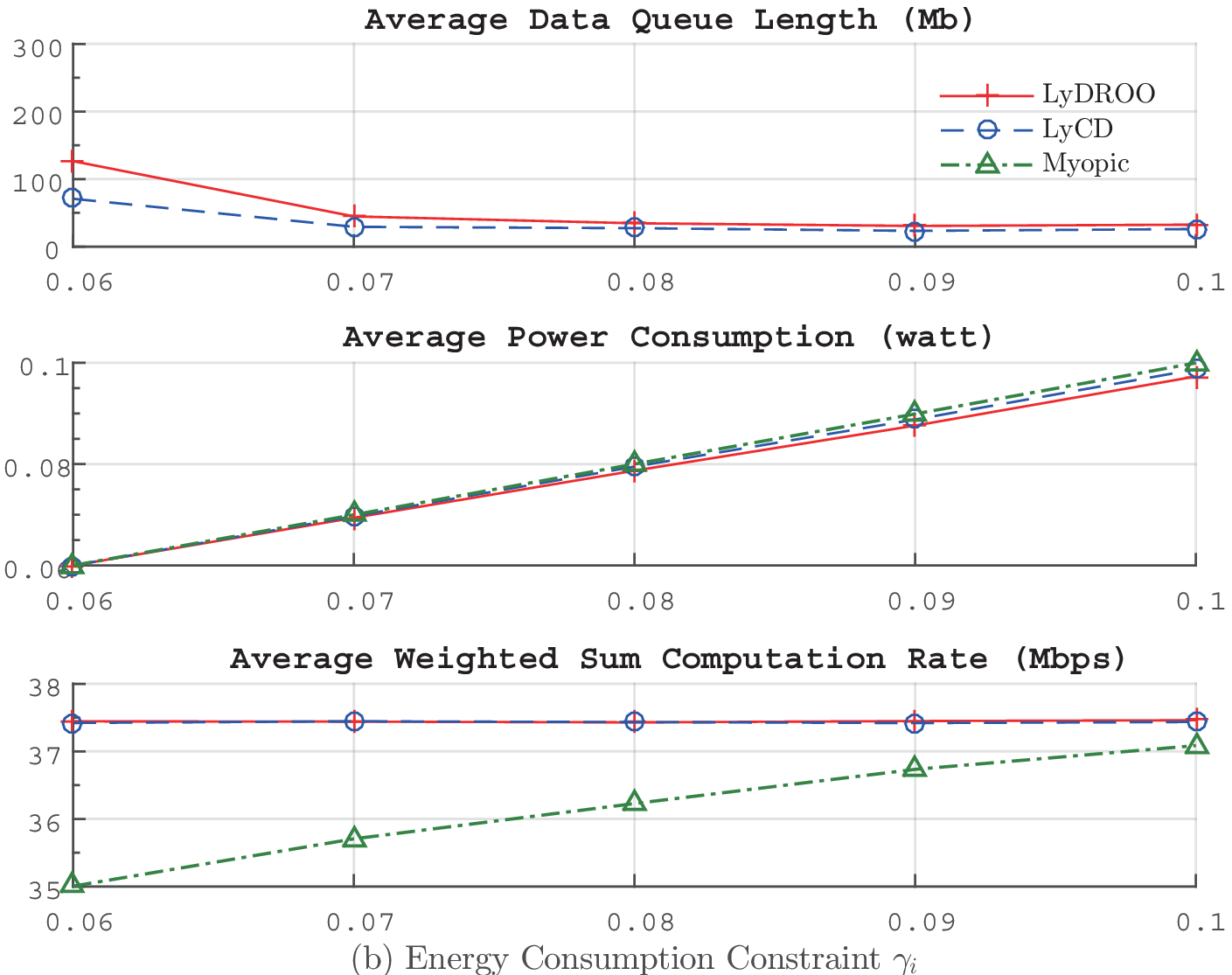}}
  \caption{Performance comparisons under different $\lambda_i$ and $\gamma_i$.}
  \label{106}
\end{figure}

In Fig.~\ref{106}, we evaluate the impact of system parameters. In Fig.~\ref{106}(a), we fix $\gamma_i=0.08$ watt and vary data arrival rate $\lambda_i$ from $2.5$ to $3.2$ Mbps. In Fig.~\ref{106}(b), we fix $\lambda_i=3$ and vary power constraint $\gamma_i$ from $0.06$ to $0.1$. We omit the results for $\lambda_i \geq 3.3$ and $\gamma_i \leq 0.05$ in the two figures, respectively, because we observe that none of the three schemes can maintain queue stability under the heavy data arrivals and stringent power constraints, i.e., arrival rate surpasses the achievable sum computation rate. All the three schemes satisfy the average power constraints under different parameters in both figures. In Fig.~~\ref{106}(a), the data queue lengths of all the three schemes increase with $\lambda_i$. In particular, the data queues are stable with LyCD and LyDROO under all the considered $\lambda_i$, while the queue lengths of the Myopic scheme become infinite when $\lambda_i \geq 2.8$. In Fig.~\ref{106}(b), the data queues are stable with LyCD and LyDROO under all the considered $\lambda_i$, and the queue length decreases with $\gamma_i$ under the less stringent power constraint. In vivid contrast, the Myopic scheme has infinite queue length under all $\lambda_i$ (thus, no point appears in the queue length figure). The results show that both LyDROO and LyCD achieve much larger stable capacity region than the Myopic method, and thus are more robust under heavy workload and stringent power constraints. We also observe that LyCD and LyDROO achieve identical computation rate performance in all the considered cases. This is because when the data queues are long-term stable, the average computation rate of the $i$th WD (departures rate of the data queue) equals the data arrival rate $\lambda_i$, and thus the achievable average weighted sum computation rate is $\sum_{i=1}^N c_i\lambda_i$ for both schemes. In fact, this also indicates that both LyDROO and LyCD achieve the optimal computation rate performance in all the considered setups. In contrast, the Myopic method achieves lower computation rate when the data queues are unstable, i.e., for $\lambda_i >2.7$ in Fig.~\ref{106}(a) and all the considered $\gamma_i \in [0.06,0.1]$ in Fig.~\ref{106}(b). Moreover, the performance gap increases under heavier workload (larger $\lambda_i$) and more stringent power constraints (smaller $\gamma_i$).

\begin{figure*}
\centering
\includegraphics[width=0.7\textwidth]{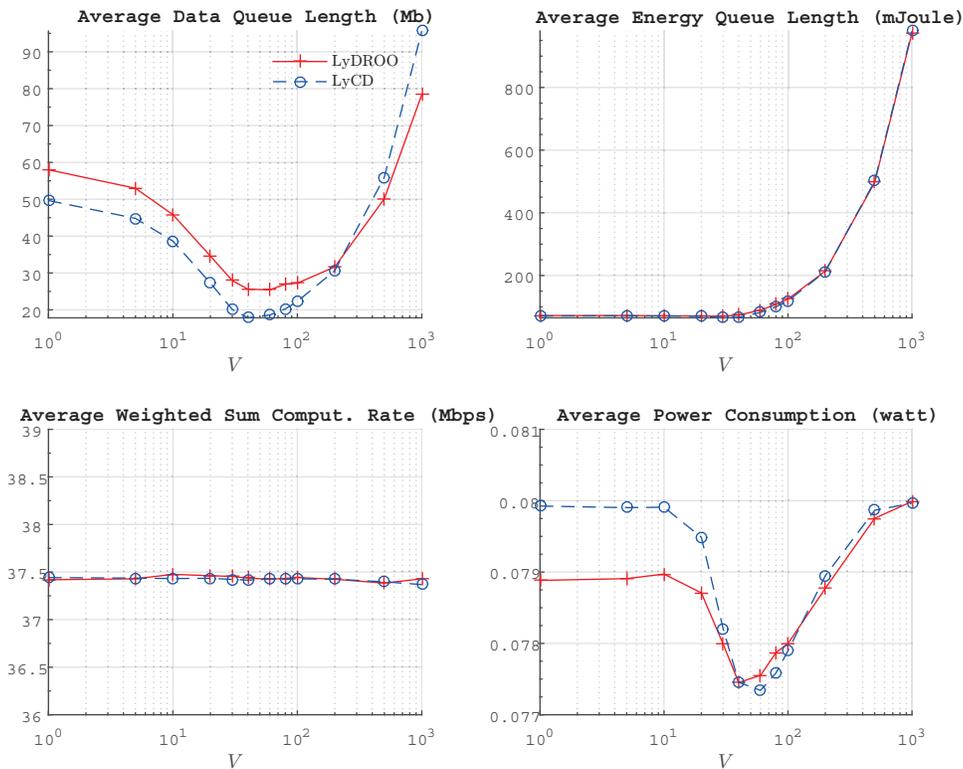}
\caption{Impact of the Lyapunov control parameter $V$.}
\label{105}
\end{figure*}

In Fig.~\ref{105}, we further show the impact of the Lyapunov control parameter $V$ on the performance of the two Lyapunov-based LyDROO and LyCD methods, where $V\in\left[1,1000\right]$. All the points in the figure are the average performance after convergence. In all the figures, the two methods achieve very similar performance, where they both maintain data and energy queues stable, control the average power consumption strictly below the threshold, and achieve the optimal computation rate performance. The parameter $V$ controls the balance between the sum computation rate performance and total data queue length. Interestingly, when $V$ is small (e.g., $V\leq 40$), the data queue length and power consumption decrease with the increase of $V$, and the virtual energy queue length is close to zero. This is because the offloading probabilities increase for most of WDs as $V$ becomes larger. However, when $V > 40$, the data queue length, power consumption, and energy queue length all increase with $V$ monotonically. This is because the offloading strategy becomes unfair when $V$ is large, such that the increase of offloading probabilities of some WDs is at the cost of decreased offloading probabilities of many others. This results in an overall increase of average data queue length and energy consumption. In practice, we should set a moderate $V$ to reduce the task data buffer size required at the WDs, which depends on the specific network deployment and the task arrival rates of all the WDs.

In Fig.~\ref{108}, we show the performance of LyDROO under different number of WDs. Specifically, we plot in Fig.~\ref{108}(a) the average queue length when the individual task arrival rate $\lambda_i$ varies. We observe that LyDROO can maintain stable task data queue for $\lambda_i\leq 3.2$ Mbps when $N=10$, $\lambda_i\leq 2.4$ Mbps when $N=20$, and $\lambda_i\leq 2$ Mbps when $N=30$. The points where task data queue becomes unstable are not plotted, e.g., $\lambda_i \geq 2.5$ Mbps for $N=20$. As expected, the stable capacity region shrinks with $N$ because of the heavier computation workload in the system under the same $\lambda_i$. For a specific individual task arrival rate $\lambda_i$, the average data queue length increases with $N$. For instance, for $\lambda_i = 2$ Mbps, the queue length is less than $5$ when $N=10$, around $20$ when $N=20$, and around $50$ when $N=30$. In Fig.~\ref{108}(b), we observe that the energy consumption increases with $\lambda_i$ for all $N$, and gradually reaches the power consumption threshold $0.08$ Watt when $\lambda_i$ approaches the upper boundary of stable capacity region. The higher power consumption arises from the more stringent resource constraint when the overall network computation workload increases.

\begin{figure}
\centering
\includegraphics[width=0.48 \textwidth]{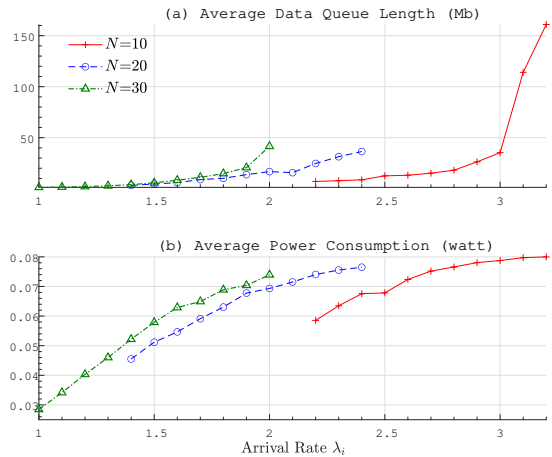}
\caption{Performance of LyDROO under different number of WDs $N\in\{10,20,30\}$. }
\label{108}
\end{figure}

From the above discussions, both LyDROO and LyCD achieve excellent computation performance under different parameters. In Table~\ref{tab:CPU}, we compare their computation time under different number of WDs $N$. Here, we consider a fixed total network workload $30$ Mbps and equally allocate $\lambda_i = 30/N$ to each WD for $N\in\left\{10,20,30\right\}$. The locations of the $N$ WDs are evenly spaced within $[120,255]$ meters distance to the ES. We observe that the two methods achieve similar computation rate performance for all $N$ and all the long-term constraints are satisfied. Besides, thanks to setting a time-varying $M_t$ in (\ref{108}), LyDROO achieves significant saving in execution time compared to that when a fixed $M_t = 2N$ is used, e.g., saves more than $80\%$ execution time for $N =30$, without degrading the convergence. Due to the page limit, we omit the illustrations of detailed performance and focus on comparing the computation time between LyCD and LyDROO methods. In Table~\ref{tab:CPU}, LyDROO takes at most $0.156$ second to generate an offloading action in all the cases. In contrast, LyCD consumes acceptable latency when $N=10$, but significantly long latency when $N=30$, e.g., around $50$ times longer than that of LyDROO method. Because the channel coherence time of a common indoor IoT system is no larger than several seconds, the long computation time makes LyCD costly even infeasible in a practical MEC system with online offloading decision. The proposed LyDROO algorithm, in contract, incurs very short latency overhead, e.g., around $3\%$ overhead when the time frame is $5$ seconds for $N=30$. Recall that after the DNN generating a control action in a time frame, the training process of the DNN is performed in parallel with task offloading and computation in the remainder of the time frame, and thus does not incur additional delay overhead. Therefore, the LyDROO algorithm can be efficiently applied in an MEC system under fast channel variation.

\begin{table}
\caption{Computation rate and CPU computation time when $N$ varies.}
\footnotesize
\begin{center}
\begin{tabular}{|c| c| c| c|c|c|}
\hline
  & \multicolumn{2}{c}{Computation rate (Mbps)} & \multicolumn{3}{|c|}{CPU computation time (second)}\\ \hline
  $N$                & LyDROO        &  LyCD       &   LyDROO      &   LyCD          &   $\frac{\text{LyCD}}{\text{LyDROO}}$\\
  $10$               & 37.43         & 37.43       &  $ 0.021$       &   $0.27$      &   12.86         \\
  $20$               & 37.61         & 37.60       &  $ 0.108$      &   $2.57$       &    23.80        \\
  $30$               & 37.36         & 37.36       &  $ 0.156$       &   $8.02$       &   51.41         \\
  \hline
\end{tabular}
\end{center}
\label{tab:CPU}
\end{table}

\section{Conclusions and Discussions}
In this paper, we have studied an online stable computation offloading problem in a multi-user MEC network under stochastic wireless channel and task data arrivals. We formulate a multi-stage stochastic MINLP problem that maximizes the average weighted sum computation rate of all the WDs under long-term queue stability and average power constraints. The online design requires making joint action of binary computation offloading and resource allocation in each short time frame without the assumption of knowing the future realizations of random channel conditions and data arrivals. To tackle the problem, we proposed a LyDROO framework that combines the advantages of Lyapunov optimization and DRL. We show in both theory and simulations that the proposed approach can achieve optimal computation rate performance meanwhile satisfying all the long-term constraints. Besides, it incurs very low computational complexity in generating an online action, and converges within relatively small number of iterations. The proposed LyDROO framework has wide application in MEC networks in enhancing both the efficiency and robustness of computation performance.

We conclude the paper with some potential extensions of the proposed LyDROO scheme and future working directions. First, besides binary computation offloading considered in this paper, the proposed LyDROO scheme can also be extended to design online partial computation offloading strategy where the computation tasks consists of multiple independent subtasks (such as in \cite{2020:Xiao}). By carefully setting binary variables to represent which subset of subtasks to be offloaded for edge execution, LyDROO is applicable to jointly optimize the binary offloading decisions and continuous resource allocation for the partial offloading scheme.

\begin{figure}
\centering
\includegraphics[width=0.48 \textwidth]{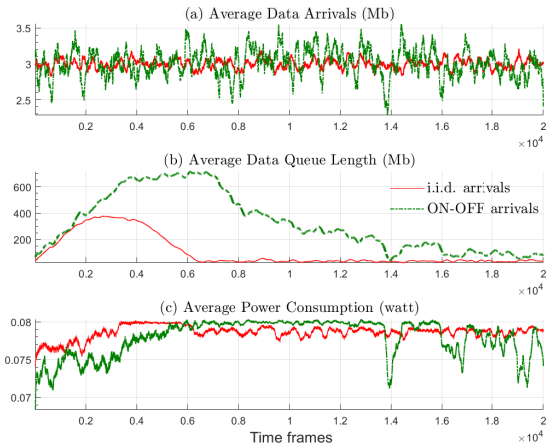}
\caption{The convergence performance of LyDROO under both i.i.d. exponential and the non-i.i.d. Markov modulated ON-OFF task arrival models.}
\label{107}
\end{figure}

Second, we consider in this paper that the task data arrivals follow an i.i.d. process, which is a crucial assumption for the proof of the convergence performance of the LyDROO scheme in Algorithm $1$. However, according to Theorem 4.9 in \cite{2010:Neely}, the proposed LyDROO can achieve the similar $\left[O(1/V),O(V)\right]$ performance guarantees as those described in Theorem $1$ of this paper when the task data arrivals follow a more general ergodic (possibly non-i.i.d.) process, such as a Markov modulated process that the distribution of arrival rates is time-varying and correlated in time. In Fig.~\ref{107}, we evaluate the performance of LyDROO under non-i.i.d. task arrivals for $N=10$ WDs, where the task arrivals follow an ON-OFF Markov modulated random process. Specifically, we consider two states for the arrival process, i.e., the ON state and the OFF state, which are modulated by a two-state Markov chain with transition matrix $[0.1, 0.9;0.9, 0.1]$. The arrived task data size $A_i^t$ at the $i$th WD in the $t$th time frame is $0$ if the system is in OFF state, and follows an i.i.d. exponential distribution in ON state. In practice, the ON-OFF Markov modulated random process models the bursty arrivals of task data. We compare the convergence performance of LyDROO under both i.i.d. exponential and the non-i.i.d. ON-OFF task arrival models. For fair comparison, we set equal long-term average task arrivals rate $3$ Mbps for both data arrival models. We plot in Fig.~\ref{107}(a) the average task arrival of the $10$ WDs over different time frames of both i.i.d. and the non-i.i.d. ON-OFF task arrival models. We observe in Fig.~\ref{107}(b) that LyDROO can achieve stable task data queue, and in fact very low task queue length, for both i.i.d. and non-i.i.d. task data arrivals after sufficient training, although the time until convergence is longer under the non-i.i.d. arrivals. In Fig.~\ref{107}(c), the average energy consumption constraint $0.08$ watt is also satisfied under both task arrival models. The results demonstrate the effectiveness of the proposed LyDROO under non-i.i.d. task data arrivals.

Third, we assume a block fading channel model in this paper. In practice, however, wireless channel may experience small variations within a time frame. Recall that $h_i^t$ denotes the channel gain at the beginning of the $t$th time frame. In case of small channel variation, we can include a signal-to-noise (SNR) power margin $\rho\geq 1$ when setting the computation offloading rate, i.e., $D^{t}_{i,O}=\frac{W\tau^t_i T}{v_u}\log_2\left(1+\frac{E_{i,O}^t h_i^t}{\tau^t_i T \rho N_0}\right)$ in (2), such that the channel gain is likely above $h_i^t/\rho$ throughout the time frame. Evidently, setting a larger $\rho$ increases the robustness of communication against channel variation, however, at the cost of lower spectrum efficiency.

Fourth, we neglect in this paper the delay on downloading the computation result from the edge server. When the downloading time is non-negligible for some application, we denote the delay on downloading the result of the $i$th offloading WD in the $t$th time frame as
\begin{equation}
      \label{3}
      w^t_i = \frac{L_i v_u}{W\log_2 \left(1+ \frac{P_0 g^t_i}{N_0}\right)},\ \forall i \in \mathcal{M}_1^t,
\end{equation}
where $g^t_i$ denotes the downlink channel gain, $P_0$ denotes the fixed transmit power of the edge base station, and $L_i$ denotes the fixed size of computation result. During the execution of the LyDROO algorithm, $\mathcal{M}_1^t$ is the output of the actor module, such that $w_i^t$'s are fixed parameters when the critic module solves the optimal resource allocation problem (P4) given $\mathcal{M}_1^t$. Therefore, we can include result downloading delay into consideration by simply replacing the time allocation constraint in (P4) $\sum_{i\in \mathcal{M}_1} \tau_i \leq 1$ with the similar linear constraint $\sum_{i\in \mathcal{M}_1^t} \left(\tau_i^t + w_i^t\right) \leq 1$, without affecting the overall algorithm design of LyDROO.

Last but not the least, in this paper, we coordinate the computation offloading of multiple WDs using TDMA. In fact, the proposed LyDROO is also applicable to MEC systems using other multiple access methods, such as FDMA, CDMA, OFDMA, and NOMA (non-orthogonal multiple access), as long as the critic module can quickly obtain the optimal wireless resource allocation. Accordingly, the technical challenge lies in the design of efficient resource allocation algorithms under different multiple access schemes considered.

\begin{appendices}

\section{Proof of Proposition $1$}
\emph{Proof}: Given $r_{i,O}$, we denote the objective of the problem (\ref{20}) as $\Omega(\tau_i)$, which is a strictly concave function within the feasible set $\tau_i \geq \frac{r_{i,O}}{R_i^{max}}$. Accordingly, the minimum is achieved at either the boundary point $\frac{r_{i,O}}{R_i^{max}}$ or the point $v_1$ that satisfies $\Omega'(v_1)=0$, depending on the value of $v_1$. To obtain $v_1$, we take the derivative of $\Omega(\tau_i)$ and set it equal to zero, i.e.,
\begin{equation}
\label{19}
\begin{aligned}
&\Omega'(\tau_i) \\
=& -\mu - \frac{Y_i(t) N_0}{h_i} \left(2^{\frac{r_{i,O} v_u}{W \tau_i}}-1 - \ln 2 \cdot 2^{\frac{r_{i,O}  v_u}{W \tau_i}} \cdot \frac{r_{i,O} v_u}{W \tau_i} \right)\\
=& -\frac{Y_i(t)N_0 e}{h_i} \bigg[ e^{-1}\left(\frac{\mu h_i}{Y_i(t)N_0} -1\right) \\
&- e^{\ln 2 \frac{r_{i,O} v_u}{W \tau_i}- 1}\left(\ln 2 \cdot\frac{r_{i,O} v_u}{W \tau_i} - 1\right)\bigg]=0,\\
\Rightarrow &  e^{\ln 2 \frac{r_{i,O} v_u}{W \tau_i}- 1}\left(\ln 2 \cdot\frac{r_{i,O} v_u}{W \tau_i} - 1\right) = e^{-1}\left(\frac{\mu h_i}{Y_i(t)N_0} -1\right).
\end{aligned}
\end{equation}
Because $e^{-1}\left(\frac{\mu h_i}{Y_i(t)N_0} -1\right)\geq -1$, the above equality is equivalent to
\begin{equation}
\ln 2 \cdot\frac{r_{i,O} v_u}{W \tau_i} - 1 = \mathcal{W}\left(e^{-1}\left[\frac{\mu h_i}{Y_i(t)N_0} -1\right]\right),
\end{equation}
where $\mathcal{W}(x)$ denotes the Lambert-W function. Therefore, we have
\begin{equation}
v_1 = \frac{\ln 2 v_u\cdot r_{i,O}}{W\cdot \left[  \mathcal{W}\left(e^{-1}\left[\frac{\mu h_i}{Y_i(t)N_0} -1\right]\right)+ 1\right]}.
\end{equation}

If $v_1 < \frac{r_{i,O}}{R_i^{max}}$, or equivalently $\Omega'(\tau_i)=0$ is not achievable within the feasible set, we can infer that the optimal solution is obtained at the boundary $\left(\tau_i\right)^* = \frac{r_{i,O}}{R_i^{max}}$. Because $\Omega(\tau_i)$ is concave, $\Omega'(\tau_i)$ is a decreasing function. Given $\Omega'(v_1)=0$, the condition $v_1 < \frac{r_{i,O}}{R_i^{max}}$ is equivalent to $\Omega'\left(\frac{r_{i,O}}{R_i^{max}}\right)<0$. By substituting $\tau_i = \frac{r_{i,O}}{R_i^{max}}$ into (\ref{19}), we have $v_1 < \frac{r_{i,O}}{R_i^{max}}$ when
\begin{equation}
\label{21}
\begin{aligned}
&\mu + Y_i(t)P_i^{max} \left[ 1 - \ln\left(1+ d_i\right)\left(\frac{1}{d_i}+ 1\right) \right]  > 0\\
\Rightarrow& \ln\left(1+d_i\right)  \leq \left(1+ \frac{\mu}{Y_i(t)P_i^{max}}\right)\left(1- \frac{1}{1+d_i}\right) \\
\Rightarrow& \ln\left(\frac{1}{1+d_i}\right) \geq -A_i + \frac{A_i}{1+d_i},
\end{aligned}
\end{equation}
where $d_i \triangleq \frac{h_i P_i^{max}}{N_0}$ and $A_i \triangleq 1+ \frac{\mu}{Y_i(t)P_i^{max}}$. By taking a natural exponential operation at both sides of (\ref{21}), we have
\begin{equation*}
\begin{aligned}
&\exp \left(-\frac{A_i}{1+d_i}\right) \left(\frac{1}{1+d_i}\right) \geq \exp\left(-A_i\right) \\
\Rightarrow & \exp \left(-\frac{A_i}{1+d_i}\right) \left(-\frac{A_i}{1+d_i}\right) \leq  - A_i\exp\left(-A_i\right).
\end{aligned}
\end{equation*}
Because the RHS of the above inequality satisfies $ -e^{-1} \leq - A_i\exp\left(-A_i\right) \leq 0$, the inequality can be equivalently expressed as
\begin{equation}
\label{23}
-A_i/(1+d_i) \leq \mathcal{W}\left( - A_i\exp\left(-A_i\right) \right),
\end{equation}
where $\mathcal{W}\left( - A_i\exp\left(-A_i\right) \right) \in \left[-1,0\right]$. The equivalence holds because $\mathcal{W}(x)$ is an increasing function when $x\geq -1/e$. After some simple manipulation, we obtain from (\ref{23}) that the optimal solution $\left(\tau_i\right)^* = \frac{r_{i,O}}{R_i^{max}}$ when $h_i \leq \frac{N_0}{P_i^{max}}\left(\frac{A_i}{-\mathcal{W}\left( - A_i\exp\left(-A_i\right) \right)}-1\right)$. Otherwise, we conclude that $v_1 \geq \frac{r_{i,O}}{R_i^{max}}$ and $\Omega'(\tau_i)=0$ is achievable such that the optimal solution is $\tau_i^*=v_1$. $\hfill \blacksquare$

\section{Proof of Theorem $1$}
To prove Theorem $1$, we first introduce the following two lemmas.

\textbf{Lemma 1}: Suppose that (P1) is feasible and $\omega(t)$ is stationary, then for any $\delta>0$, there exits an $\omega$-only policy $\Gamma$, such that the following inequalities are satisfied:
\begin{equation}
\label{77}
\begin{aligned}
&\mathbb{E}\left[R^t\left(\alpha^{\Gamma,t}\right)\right]  \geq R^{opt} - \delta, \\
&\mathbb{E}\left[e_i^t\left(\alpha^{\Gamma,t}\right) - \gamma_i \right] \leq \delta, \ \forall i, \\
&\mathbb{E}\left[A_i^t\right] \leq \mathbb{E}\left[D_i^t\left(\alpha^{\Gamma,t}\right)\right] + \delta, \ \forall i.
\end{aligned}
\end{equation}

\emph{Proof}: See Theorem 4.5 of \cite{2010:Neely} for detailed proof. $\hfill \blacksquare$

\textbf{Lemma 2}: If $Y_i(t)$ is rate stable, i.e., $\lim_{K \rightarrow \infty} \frac{Y_i(K)}{K} =0$ holds with probability $1$, then the $i$th average power constraint in (\ref{63}) is satisfied with probability $1$.

\emph{Proof}: Using the sample path property (Lemma 2.1 of \cite{2010:Neely}), we have
\begin{equation}
\begin{aligned}
& \frac{Y_i(K)}{K} -  \frac{Y_i(1)}{K} \geq \frac{1}{K} \mathsmaller\sum_{t=1}^K e^t_i - \frac{1}{K} \mathsmaller\sum_{t=1}^K \gamma_i \\
\Rightarrow& \frac{1}{K} \mathsmaller \sum_{t=1}^K e^t_i \leq \gamma_i +  \frac{Y_i(K)}{K}.
\end{aligned}
\end{equation}
By taking the limit $K\rightarrow \infty$ on both size and substituting $\lim_{K \rightarrow \infty} \frac{Y_i(K)}{K} =0$, we have $\lim_{K\rightarrow \infty}\frac{1}{K} \sum_{t=1}^K e^t_i \leq \gamma_i$ holds with probability $1$, which completes the proof. $\hfill \blacksquare$

\emph{Proof of Theorem $1$}: Because (P1) is feasible and $\omega(t)$ is an i.i.d. process, we apply Lemma 1 and consider a fixed $\delta>0$ and the corresponding $\omega$-only control policy $\Gamma$. Because the minimum of (\ref{12}) is obtained over all feasible control policies, including $\Gamma$, we have
\begin{equation}
\label{74}
\begin{aligned}
&\Delta L\left(\mathbf{Z}(t)\right) - V \cdot \mathbb{E}\left[R^t|\mathbf{Z}(t)\right] \\
\leq&  \hat{B} + C+ \mathsmaller\sum_{i=1}^N  \Big(Q_{i}(t) \mathbb{E}\left[\left(A_{i}^t-D^{t}_{i}\left(\alpha^{\Gamma,t}\right)\right)| \mathbf{Z}(t)\right] \\
&+  Y_i(t) \mathbb{E}\left[e^t_i\left(\alpha^{\Gamma,t}\right) -\gamma_i| \mathbf{Z}(t)\right] - V \cdot \mathbb{E} \left[ R^t\left(\alpha^{\Gamma,t}\right)|\mathbf{Z}(t)\right] \Big)\\
\overset{\dag}{\leq}  &  \hat{B} + C + \mathsmaller\sum_{i=1}^N  Q_{i}(t) \mathbb{E}\left[\left(A_{i}^t-D^{t}_{i}\left(\alpha^{\Gamma,t}\right)\right)\right] \\
&+ \mathsmaller \sum_{i=1}^N  Y_i(t) \mathbb{E}\left[e^t_i\left(\alpha^{\Gamma,t}\right) -\gamma_i \right] - V \cdot \mathsmaller\sum_{i=1}^N \mathbb{E} \left[ R^t\left(\alpha^{\Gamma,t}\right) \right] \\
\overset{\ddag}{\leq}& \hat{B} +  C + \delta \left[\mathsmaller\sum_{i=1}^N  \left(Q_{i}(t) + Y_i(t) \right)\right] - V\left(R^{opt} - \delta \right),
\end{aligned}
\end{equation}
where inequality $\left(\dag\right)$ is because the control policy $\Gamma$ is independent to queue backlog $\mathbf{Z}(t)$, and the inequality $\left(\ddag\right)$ is obtained by plugging (\ref{77}). By letting $\delta\rightarrow 0$, we have
\begin{equation}
\label{75}
\begin{aligned}
\Delta L\left(\mathbf{Z}(t)\right) - V \cdot \mathbb{E}\left[R^t|\mathbf{Z}(t)\right] \leq \hat{B} +  C  - V R^{opt}.
\end{aligned}
\end{equation}
Furthermore, by summing both sizes of (\ref{75}) from $t=1$ to $K$, and taking iterated expectations and telescoping sums, then dividing both sizes by $KV$, we obtain
\begin{equation}
\label{121}
\begin{aligned}
&  \frac{1}{KV}\left(\mathbb{E} \left[L\left(\mathbf{Z}(K+1)\right)\right] - \mathbb{E} \left[L\left(\mathbf{Z}(1)\right)\right]  -  \mathsmaller \sum_{t=1}^{K} \mathbb{E}\left[R^t\right] \right)    \\
& \leq   (\hat{B} +  C)/V  -  R^{opt}.
\end{aligned}
\end{equation}
Because $L\left(\mathbf{Z}(K+1)\right)\geq 0$ and $L\left(\mathbf{Z}(1)\right) =0$, we prove a) by letting $K\rightarrow \infty$ in (\ref{121}).

To prove b), we consider the $\omega$-only policy $\Pi$ that satisfies the Slater condition for some values $\epsilon$ and $\Phi\left(\epsilon\right)$. By plugging the policy $\Pi$ to the RHS of the inequality ($\dag$) in (\ref{74}), we have
\begin{equation}
\begin{aligned}
&\Delta L\left(\mathbf{Z}(t)\right) - V \cdot \mathbb{E}\left[R^t|\mathbf{Z}(t)\right] \\
&\leq \hat{B} +  C - \epsilon \left[\mathsmaller\sum_{i=1}^N  \left(Q_{i}(t) + Y_i(t) \right)\right] - V \Phi\left(\epsilon\right),
\end{aligned}
\end{equation}
where the inequality is obtained from (\ref{81}). Taking iterated expectations, summing the telescoping series, and rearranging terms yields
\begin{equation*}
\begin{aligned}
&1/K \mathsmaller \sum_{t=1}^{K} \mathsmaller\sum_{i=1}^N  \mathbb{E}\left[\left(Q_{i}(t) + Y_i(t) \right)\right] \\
&\leq \frac{\hat{B} +  C  + V\left( \frac{1}{K} \cdot \mathsmaller \sum_{t=1}^{K}  \mathbb{E}\left[R^t\right]-\Phi\left(\epsilon\right)\right)}{\epsilon} + \frac{\mathbb{E} \left[L\left(\mathbf{Z}(1)\right)\right]}{\epsilon K}.
\end{aligned}
\end{equation*}
By letting $K\rightarrow \infty$ and plugging the fact that $\lim_{K\rightarrow \infty} \frac{1}{K} \sum_{t=1}^{K}  \mathbb{E}\left[R^t\right] \leq  R^{opt}$, we have
\begin{equation}
\label{84}
\begin{aligned}
&\lim_{K\rightarrow \infty} \frac{1}{K}\mathsmaller\sum_{t=0}^{K-1} \mathsmaller\sum_{i=1}^N  \mathbb{E}\left[\left(Q_{i}(t) + Y_i(t) \right)\right] \\
&\leq \frac{\hat{B} +  C  + V\left( R^{opt}-\Phi\left(\epsilon\right)\right)}{\epsilon}.
\end{aligned}
\end{equation}
Then, (\ref{76}) in b) is proved because $Y_i(t)\geq 0$. Meanwhile, (\ref{84}) also indicates that
\begin{equation}
\begin{aligned}
& \lim_{K\rightarrow \infty} \frac{1}{K} \mathsmaller\sum_{t=1}^K \mathbb{E}\left[Q_i(t)\right] <\infty, \ \forall i,\\
& \lim_{K\rightarrow \infty} \frac{1}{K} \mathsmaller\sum_{t=1}^K \mathbb{E}\left[Y_i(t)\right] <\infty,\ \forall i.
\end{aligned}
\end{equation}
That is, all the data queues and virtual queues are strongly stable. Because strong stability implies rate stability (Theorem 2.8 of \cite{2010:Neely}), we have $Y_i(t)$ is rate stable. By Lemma 2, the average power constraint (\ref{63}) is satisfied with probability $1$, which leads to the proof of c). $\hfill \blacksquare$

\end{appendices}

\end{document}